\documentclass[showpacs,preprintnumbers,superscriptaddress,twocolumn]{revtex4-1}
\usepackage{amssymb}
\usepackage{amsfonts}
\usepackage{latexsym}
\usepackage{dcolumn}
\usepackage{amsmath}
\usepackage{graphicx}
\usepackage{psfrag,pstricks}
\usepackage[normalem]{ulem}
\usepackage{float}

\begin{document}
\title{Quantum Illumination at the Microwave Wavelengths}
\author{Shabir Barzanjeh}
\affiliation{Institute for
Quantum Information, RWTH Aachen University, 52056 Aachen,
Germany}
\author{Saikat Guha}
\affiliation{Quantum Information Processing Group, Raytheon BBN
Technologies, Cambridge, Massachusetts 02138, USA}
\author{Christian Weedbrook}
\affiliation{QKD Corp., 60 St. George St., Toronto, M5S 3G4,
Canada}
\author{David Vitali}
\affiliation{School of Science and Technology, University of
Camerino, Camerino, Macerata 62032, Italy}
\author{Jeffrey H. Shapiro}
\affiliation{Research Laboratory of Electronics, Massachusetts
Institute of Technology, Cambridge, Massachusetts 02139, USA}
\author{Stefano Pirandola}
\email{stefano.pirandola@york.ac.uk} \affiliation{Department of
Computer Science \& York Centre for Quantum Technologies,
University of York, York YO10 5GH, United Kingdom}

\begin{abstract}
Quantum illumination is a quantum-optical sensing technique in which an
entangled source is exploited to improve the detection of a low-reflectivity
object that is immersed in a bright thermal background. Here we describe and
analyze a system for applying this technique at microwave frequencies, a
more appropriate spectral region for target detection than the optical, due
to the naturally-occurring bright thermal background in the microwave
regime. We use an electro-opto-mechanical converter to entangle microwave
signal and optical idler fields, with the former being sent to probe the
target region and the latter being retained at the source. The microwave
radiation collected from the target region is then phase conjugated and
upconverted into an optical field that is combined with the retained idler
in a joint-detection quantum measurement. The error probability of this
microwave quantum-illumination system, or quantum radar, is shown to be
superior to that of any classical microwave radar of equal transmitted
energy.
\end{abstract}

\maketitle


\textit{Introduction}.--~Entanglement is the foundation of many quantum
information protocols~\cite{Nielsen,Wilde,Cerf}, but it is easily destroyed
by environmental noise that, in almost all cases, kills any benefit such
nonclassical correlations would otherwise have provided. Quantum
illumination (QI)~\cite{Lloyd,pirandola}, however, is a notable exception:
it thrives in environments so noisy that they are entanglement breaking.

The original goal of QI was to detect the presence of a
low-reflectivity object, immersed in a bright thermal background,
by interrogating the target region with one optical beam while
retaining its entangled counterpart for subsequent joint
measurement with the light returned from that target region.
Although the thermal noise destroys the entanglement, theory
showed that the QI system will significantly outperform a
classical (coherent-state) system of the same transmitted
energy~\cite{pirandola,saikat,Weedbrook}. Later, a QI protocol was
proposed for secure communication~\cite{Shapiro2} whose
experimental realization~\cite{Zhang} showed that entanglement's
benefit could indeed survive an entanglement-breaking channel.
Because of this feature, QI is perhaps one of the most surprising
protocols for quantum sensing. Together with quantum
reading~\cite{Read1,Read2,Read3,Read4}, it represents a practical
example of quantum channel discrimination~\cite{Cerf}, in which
entanglement is beneficial for a technologically-driven
information task.

So far, QI has only been demonstrated at optical wavelengths~\cite%
{Zhang,Lopaeva,ZhangDetection}, for which naturally-occurring
background radiation contains far less than one photon per mode on
average, even though QI's performance advantage \emph{requires}
the presence of a bright background. The QI communication protocol
from~\cite{Shapiro2,Zhang} deals with this problem in a natural
way by purposefully injecting amplified spontaneous emission noise
to thwart eavesdropping. By contrast, similar noise injection in
QI target-detection experiments has to be considered artificial,
because better target-detection performance would be obtained
without it. The appropriate wavelengths for QI-enabled target
detection thus lie in the microwave region, in which almost all
radar systems operate and in which there is naturally-occurring
background radiation containing many photons per mode on average.
In general, the development of quantum information techniques for
microwave frequencies is quite
challenging~\cite{Ralph,Ralph2,Ottaviani}.

In this Letter, we introduce a novel QI target-detection system that
operates in the microwave regime. Its transmitter uses an
electro-opto-mechanical (EOM) converter~\cite%
{shabir2,shabir1,lenh1,bochmann,bagci} in which a mechanical resonator
entangles signal and idler fields emitted from microwave and optical
cavities~\cite{shabir2, shabir1}. Its receiver employs another EOM
device---operating as a phase-conjugator \emph{and} a wavelength
converter---whose optical output is used in a joint measurement with the
retained idler. We show that our system dramatically outperforms a
conventional (coherent-state) microwave radar of the same transmitted
energy, achieving an orders-of-magnitude lower detection-error probability.
Moreover, our system can be realized with state-of-the-art technology, and
is suited to such potential applications as standoff sensing of
low-reflectivity objects, and environmental scanning of electrical circuits.
Thanks to its enhanced sensitivity, our system could also lead to low-flux
non-invasive techniques for protein spectroscopy and biomedical imaging.

\begin{figure}[h!]
\vspace{-0.2cm} \centering
\includegraphics[width=3.5in]{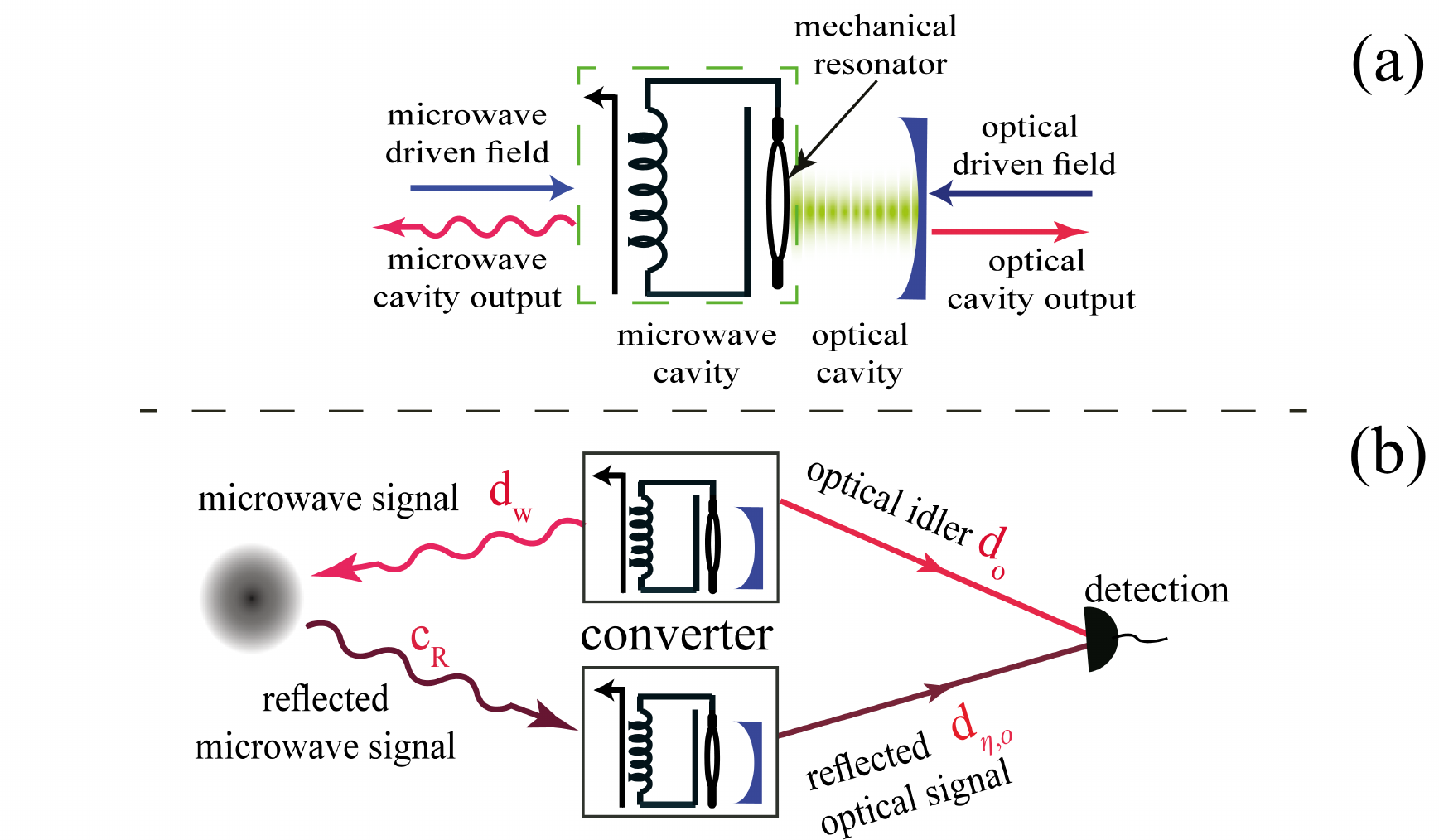}
\vspace{-0.4cm} \caption{(a) Schematic of the
electro-opto-mechanical (EOM) converter in which driven microwave
and optical cavities are coupled by a mechanical resonator. (b)
Microwave-optical QI using EOM converters. The transmitter's EOM
converter entangles microwave and optical fields. The receiver's
EOM converter transforms the returning microwave field to the
optical domain while performing a phase-conjugate operation.}
\label{fig1}
\end{figure}

\textit{Electro-opto-mechanical converter}.--~As depicted in Fig.~\ref{fig1}%
(a), the EOM converter couples a microwave-cavity mode (annihilation
operator $\hat{a}_{\text{w}}$, frequency $\omega _{\text{w}}$, damping rate $%
\kappa _{\text{w}}$) to an optical-cavity mode (operator $\hat{a}_{o}$,
frequency $\omega _{o}$, damping rate $\kappa _{o}$) through a mechanical
resonator (operator $\hat{b}$, frequency $\omega _{M}$, damping rate $\gamma
_{M}$) \cite{shabir1,lenh1, bochmann}. In the frame rotating at the
frequencies of the microwave and optical driving fields, the interaction
between the cavities' photons and the resonator's phonons is governed by the
Hamiltonian \cite{supp}
\begin{equation*}
\hat{H}=\hbar \omega _{M}\hat{b}^{\dagger }\hat{b}+\hbar \sum_{j=\text{w},o}%
\left[ \Delta _{0,j}+g_{j}(\hat{b}+\hat{b}^{\dagger })\right] \hat{a}%
_{j}^{\dagger }\hat{a}_{j}+\hat{H}_{\text{dri}}.
\end{equation*}%
Here, $g_{j}$ is the coupling rate between the mechanical resonator and
cavity $j$, which is driven at frequency $\omega _{j}-\Delta _{0,j}$ by the
coherent-driving Hamiltonian $\hat{H}_{\text{dri}}$~\cite{supp}.

The electro-opto-mechanical coupling rates $g_j$ are typically small, so
that we can linearize the Hamiltonian by expanding the cavity modes around
their steady-state field amplitudes $\hat{c}_{j}=\hat{a}_{j}-\sqrt{N_{j}}$,
where the $N_{j}\gg1$ are the mean numbers of cavity photons induced by the
pumps~\cite{shabir2, lenh}. In the interaction picture with respect to the
free Hamiltonian, we may then write~\cite{supp}%
\begin{equation}
\hat{H}=\hbar G_{o}(\hat{c}_{o}\hat{b}+\hat{b}^{\dagger}\hat{c}_{o}^{\dagger
})+\hbar G_{\mathrm{w}}(\hat{c}_{\mathrm{w}}\hat{b}^{\dagger}+\hat{b}\hat {c}%
_{\mathrm{w}}^{\dagger}),  \label{hameffMainText}
\end{equation}
where $G_{j}=g_{j}\sqrt{N_j}$ is the multi-photon coupling rate. This
expression assumes that the effective cavity detunings satisfy $\Delta_{%
\mathrm{w}}=-\Delta_{o}=\omega_{M}$ and that resonator is in its
fast-oscillation regime, so that the red sideband drives the microwave
cavity while the blue sideband drives the optical cavity and we can neglect
terms oscillating at $\pm 2\omega_{M}$.

Equation~(\ref{hameffMainText}) shows that the mechanical
resonator mediates a delayed interaction between the optical and
microwave cavity modes. Its first term is a parametric
down-conversion interaction that entangles the mechanical
resonator and the optical cavity mode. This entanglement is
transmitted to the propagating optical mode $\hat{d}_{o}$, if the
opto-mechanical rate $G_{o}^{2}/\kappa _{o}$ exceeds the
decoherence rate of the mechanical resonator
$r=\gamma_{M}\bar{n}_{b}^{T}\approx
\gamma_{M}k_{B}T_{\text{EOM}}/\hbar \omega _{M}$, where $k_{B}$ is
Boltzmann's constant, $T_{\text{EOM}}$ is the EOM converter's
absolute temperature, $\bar{n}_{b}^{T}=[\mathrm{e}^{\hbar \omega
_{M}/(k_{B}T_{\text{EOM}})}-1]^{-1}$, and the approximation
presumes $k_{B}T_{\text{EOM}}\ \gg \hbar \omega _{M}$, as will
be the case for the parameter values assumed later. The second term in Eq.~(%
\ref{hameffMainText}) is a beam-splitter interaction between the
mechanical resonator and the microwave cavity mode that transfers
the entanglement to the propagating microwave field
$\hat{d}_{\text{w}}$, as long as the
microwave-mechanical rate satisfies $G_{\mathrm{w}}^{2}/\kappa _{\mathrm{w}%
}>r$~\cite{lenh,hofer}.

\textit{Microwave-optical entanglement}.--~The output propagating modes can
be expressed in terms of the intracavity quantum noise operators, $\hat{c}%
_{j,\mathrm{in}}$, and the quantum Brownian noise operator,
$\hat{b}_{\mathrm{int}}$, via~\cite{supp}
\begin{align}
\hat{d}_{\mathrm{w}}& =A_{\mathrm{w}}\hat{c}_{\mathrm{w},\mathrm{in}}-B\hat{c%
}_{o,\mathrm{in}}^{\dagger }-C_{\mathrm{w}}\hat{b}_{\text{int}},
\label{qle1MAIN} \\
\hat{d}_{o}& =B\hat{c}_{{\mathrm{w}},\mathrm{in}}^{\dagger }+A_{o}\hat{c}_{o,%
\mathrm{in}}-C_{o}\hat{b}_{\text{int}}^{\dagger },  \label{downconverter2}
\end{align}%
where $A_{j}$, $B$, and $C_{j}$ depend on the cooperativity terms $\Gamma
_{j}=G_{j}^{2}/\kappa _{j}\gamma _{M}$ as given in~\cite{supp}. The $\hat{c}%
_{\mathrm{w},\mathrm{in}},\hat{c}_{\mathrm{o,in}}$ and
$\hat{b}_{\text{int}}$ modes in Eqs.~(\ref{qle1MAIN}) and
(\ref{downconverter2}) are in independent
thermal states whose average photon numbers, $\bar{n}_{\text{w}}^{T}$, $\bar{%
n}_{o}^{T}$, and $\bar{n}_{b}^{T}$, are given by temperature-$T_{\text{EOM}}$
Planck laws at their respective frequencies. It follows that the propagating
modes, $\hat{d}_{\text{w}}$ and $\hat{d}_{o}$, are in a zero-mean,
jointly-Gaussian state completely characterized by the second moments,
\begin{align*}
\bar{n}_{\text{w}}& \equiv \langle \hat{d}_{\text{w}}^{\dagger }\hat{d}_{%
\text{w}}\rangle =|A_{\text{w}}|^{2}\bar{n}_{\text{w}}^{T}+|B|^{2}(\bar{n}%
_{o}^{T}+1)+|C_{\text{w}}|^{2}\bar{n}_{b}^{T}, \\
\bar{n}_{o}& \equiv \langle \hat{d}_{o}^{\dagger }\hat{d}_{o}\rangle
=|B|^{2}(\bar{n}_{\text{w}}^{T}+1)+|A_{o}|^{2}\bar{n}_{o}^{T}+|C_{o}|^{2}(%
\bar{n}_{b}^{T}+1), \\
\langle \hat{d}_{\text{w}}\hat{d}_{o}\rangle & =A_{\text{w}}B(\bar{n}_{\text{%
w}}^{T}+1)-BA_{o}\bar{n}_{o}^{T}+C_{\text{w}}C_{o}(\bar{n}_{b}^{T}+1).
\end{align*}

\begin{figure}[h]
\vspace{-0.1cm} \centering
\includegraphics[width=3.52in]{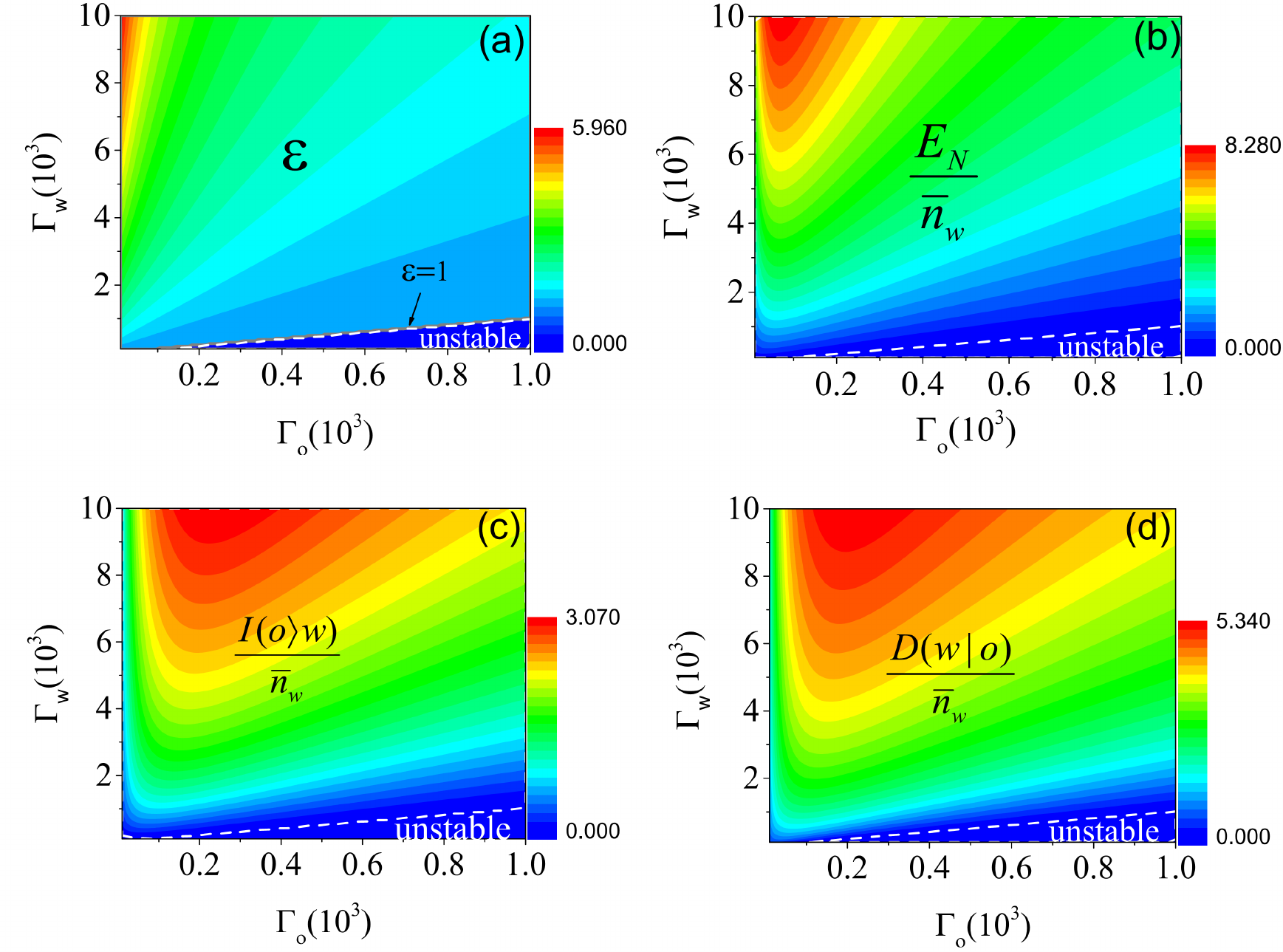}
\vspace{-0.5cm} \caption{Performance of our microwave-optical
source versus the cooperativity parameters $\Gamma _{\text{w}}$
and $\Gamma _{o}$. We show the behaviour of the entanglement
metric $\mathcal{E}$ (abstract units) in panel (a), the normalized
logarithmic negativity $E_{N}/\bar{n}_{\text{w}}$ (ebits
per microwave photon) in panel (b), the normalized coherent information $%
I(o\rangle $w$)/\bar{n}_{\text{w}}$ (qubits per microwave photon) in panel
(c), and the normalized quantum discord $D({\text{w}}|o)/\bar{n}_{\text{w}}$
(discordant bits per microwave photon) in panel (d). In all panels we assume
experimentally-achievable parameters~\protect\cite{lenh,palo}: a 10-ng-mass
mechanical resonator with $\protect\omega _{M}/2\protect\pi =10$\thinspace
MHz and $Q=30\times 10^{3}$; a microwave cavity with $\protect\omega _{%
\mathrm{w}}/2\protect\pi =10$\thinspace GHz and $\protect\kappa _{\mathrm{w}%
}=0.2\,\protect\omega _{M}$; and a 1-mm-long optical cavity with $\protect%
\kappa _{o}=0.1\,\protect\omega _{M}$ driven by a 1064-nm-wavelength laser.
The opto-mechanical and electro-mechanical coupling rates are $g_{\mathrm{o}%
}/2\protect\pi =115.512$\thinspace Hz and $g_{\mathrm{w}}/2\protect\pi %
=0.327 $\thinspace Hz, and the entire EOM converter is held at temperature $%
T_{\text{EOM}}\ =30$\thinspace mK. In each panel, the boundary between
stable and unstable operation was obtained from the Routh-Hurwitz criterion~%
\protect\cite{Stability}.}
\label{fig2}
\end{figure}

The propagating microwave and optical fields will be entangled if and only
if the metric $\mathcal{E}\equiv |\langle \hat{d}_{\text{w}}\hat{d}%
_{o}\rangle |/\sqrt{\bar{n}_{\text{w}}\bar{n}_{o}}$ is greater than $1$~\cite%
{supp}. As we can see from Fig.~\ref{fig2}, there is a wide region where $%
\mathcal{E}>1$ in the plane of the cooperativity parameters, $\Gamma _{\text{%
w}}$ and $\Gamma _{o}$, varied by varying the microwave and
optical powers driving their respective cavities, and assuming
experimentally-achievable system parameters~\cite{lenh, palo}. The
threshold condition $\mathcal{E}=1$ almost coincides with the
boundary between the stable and unstable parameter regions, as
given by the Routh-Hurwitz criterion~\cite{Stability}.

The quality of our microwave-optical source can also be evaluated using
measures of quantum correlations, as typical in quantum information. Since
the QI's advantage is computed at fixed mean number of microwave photons $%
\bar{n}_{\text{w}}$ irradiated through the target, the most powerful quantum
resources are expected to be those maximizing their quantum correlations per
microwave photon emitted. Following this physical intuition, we analyze our
source in terms of the normalized log-negativity$~$\cite{eis} $E_{N}/\bar{n}%
_{\text{w}}$ and the normalized coherent information~\cite{coh1,coh2} $%
I(o\rangle $w$)/\bar{n}_{\text{w}}$. Respectively, they represent an upper
and a lower bound to the mean number of entanglement bits (ebits) which are
distillable for each microwave photon emitted by the source~\cite{supp}.
Furthermore, since our source is in a mixed state (more precisely, a
two-mode squeezed thermal state), we also quantify its normalized quantum
discord~\cite{Discord,supp} $D({\text{w}}|o)/\bar{n}_{\text{w}}$, which
captures the quantum correlations carried by each microwave photon. As we
can see from Fig.~\ref{fig2}, our source has a remarkable performance in
producing distillable ebits and discordant bits for each microwave photon
emitted.

\textit{Microwave quantum illumination}.--~For QI target
detection, our signal-idler mode pair analysis must be extended to
a continuous-wave EOM converter whose
$W_m$-Hz-bandwidth~\cite{bandwidth} output fields are used in a
$t_m$-sec-duration measurement involving $M = t_mW_m \gg 1$
independent, identically-distributed (iid) mode pairs. The $M$
signal modes
interrogate the target region that is equally likely to contain (hypothesis $%
H_1$) or not contain (hypothesis $H_0$) a low-reflectivity object. Either
way, the microwave field that is returned consists of $M$ iid modes. Using $%
\hat{c}_R$ to denote the annihilation operator for the mode returned from
transmission of $\hat{d}_\text{w}$, we have that $\hat{c}_R = \hat{c}_B$
under hypothesis $H_0$, and $\hat{c}_R = \sqrt{\eta}\,\hat{d}_\text{w} +
\sqrt{1-\eta}\,\hat{c}_B$, under hypothesis $H_1$. Here, $0 < \eta \ll 1$ is
the roundtrip transmitter-to-target-to-receiver transmissivity (including
propagation losses and target reflectivity), and the background-noise mode, $%
\hat{c}_B$, is in a thermal state with temperature-$T_B$ Planck-law average
photon number $\bar{n}_{B}$ under $H_0$, and in a thermal state with $\bar{n}%
_{B}/(1-\eta) \approx \bar{n}_B$ under $H_1$~\cite{pirandola}. See
Fig.~\ref{fig1}(b).

Under $H_{1}$, the returned microwave and the retained optical fields are in
a zero-mean, jointly-Gaussian state with a nonzero phase-sensitive cross
correlation $\langle \hat{c}_{R}\hat{d}_{o}\rangle $ that is invariant to
the $\bar{n}_{B}$ value, while $\langle \hat{c}_{R}^{\dagger }\hat{c}%
_{R}\rangle $ increases with increasing $\bar{n}_{B}$. Consequently, the
returned and retained radiation under $H_{1}$ will \emph{not} be entangled
when
\begin{equation*}
\bar{n}_{B}\geq \bar{n}_{B}^{\text{thresh}}\equiv \eta \left( |\langle \hat{d%
}_{\text{w}}\hat{d}_{o}\rangle |^{2}/\bar{n}_{o}-\bar{n}_{\text{w}}\right) .
\end{equation*}

\textit{Microwave-to-optical phase-conjugate receiver}.--~The receiver
passes the $M$ return modes into the microwave cavity of another (identical)
EOM converter to produce $M$ iid optical-output modes each given by $\hat{d}%
_{\eta ,\mathrm{o}}=B\hat{c}_{R}^{\dagger }+A_{\mathrm{o}}\hat{c}_{\mathrm{%
o,in}}^{\prime }-C_{\mathrm{o}}\hat{b}_{\text{int}}^{^{\prime }\dagger }$,
where $\{\hat{c}_{\text{w,in}}^{\prime },\hat{c}_{o,\text{in}}^{\prime },%
\hat{b}_{\text{int}}^{\prime }\}$ have the same states as their counterparts
in the transmitter's EOM converter. The receiver's EOM converter thus phase
conjugates the returned microwave field \emph{and} upconverts it to an
optical field. This output is combined with the retained idler on a 50--50
beam splitter whose outputs are photodetected and their photon counts---over
the $t_{m}$-sec-long measurement interval---are subtracted to yield an
outcome from which a minimum error-probability decision about object absence
or presence will be made~\cite{saikat}. For $M\gg 1$, the resulting error
probability is~\cite{supp,saikat} $P_{\text{QI}}^{(M)}=\mathrm{erfc}\left(
\sqrt{\text{SNR}_{\text{QI}}^{(M)}/8}\,\right) /2,$ with SNR$_{\text{QI}%
}^{(M)}$ being the QI system's signal-to-noise ratio for its $M$
mode pairs~\cite{supp} and erfc(...) being the complementary error
function~\cite{saikat}.

\textit{Comparison with classical microwave transmitters}.-- Suppose that a
coherent-state microwave transmitter---emitting $M\bar{n}_{\text{w}}$
photons on average, with $\bar{n}_{\text{w}}$ equaling the mean number of
microwave photons per mode emitted by our source---is used to assess target
absence or presence. Homodyne detection of the microwave field returned from
the target region followed by minimum error-probability processing of its
output results in an error probability~\cite{saikat} $P_{\text{coh}}^{(M)}=%
\mathrm{erfc}\!\left( \sqrt{\text{SNR}_{\text{coh}}^{(M)}/8}\,\right) \!/2$,
in terms of this system's signal-to-noise ratio, $\text{SNR}_{\text{coh}%
}^{(M)}=4\eta M\bar{n}_{\text{w}}/(2\bar{n}_{B}+1)$. This performance
approximates the error exponent of the quantum Chernoff bound~\cite%
{QCB1,QCB2,QCBgauss} computed for $M\gg 1$, implying that homodyne detection
is the asymptotically optimal receiver for target detection when a
coherent-state transmitter is employed.

\begin{figure}[h!]
\vspace{-0.1cm} \centering
\includegraphics[width=2.5in]{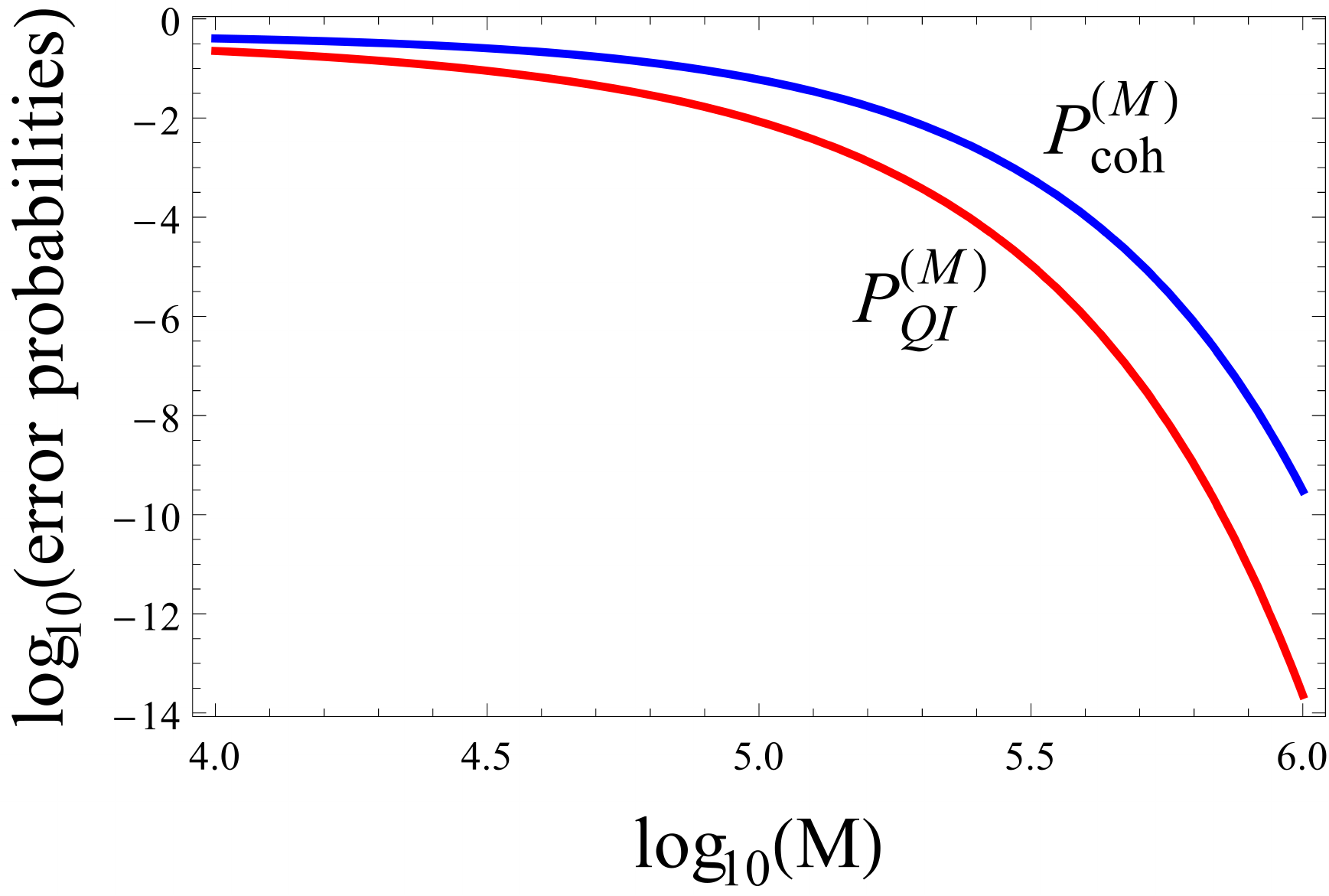}
\vspace{-0.1cm} \caption{$P_{\text{QI}}^{(M)}$ and
$P_{\text{coh}}^{(M)}$\ versus the
time-bandwidth product, $M$, assuming $\protect\eta =0.07$, $\Gamma _{\text{w%
}}\ =5181.95$ and $\Gamma _{o}=668.43$ (implying $\bar{n}_{\text{w}}\ =0.739$
and $\bar{n}_{o}=0.681$), and room temperature $T_{B}=293\,$K (implying $%
\bar{n}_{B}=610\gg \bar{n}_{B}^{\text{thresh}}=0.069$).}
\label{QCB}
\end{figure}

Figure~\ref{QCB} plots $P_{\text{QI}}^{(M)}$ and $P_{\text{coh}}^{(M)}$
versus $\log _{10}M$ for the EOM converter parameters given in Fig.~\ref%
{fig2} and $\eta =0.07$. It assumes $\Gamma _{\text{w}}\ =5181.95$ and $%
\Gamma _{o}=668.43$ (implying $\bar{n}_{\text{w}}=0.739$ and $\bar{n}%
_{o}=0.681$) and $T_{B}=293\,$K (implying $\bar{n}_{B}=610$). We
see that the QI system can have an error probability that is
orders of magnitude lower than that of the coherent-state system.
Moreover, according to Ref.~\cite{pirandola}, no other
classical-state system with the same energy constraint can have a
lower error probability than the coherent-state system.

To further study the performance gain of our microwave QI system over a
classical sensor, we evaluate $\mathcal{F}\equiv \text{SNR}_{\text{QI}%
}^{(M)}/\text{SNR}_{\text{coh}}^{(M)}$ for large $M$. This figure of merit
depends on the cooperativity parameters, $\Gamma _{\text{w}}$ and $\Gamma
_{o}$, whose values are typically large $\Gamma _{j}\gg 1$ (cf.\ the values
in Fig.~\ref{fig2}, which rely on experimentally-achievable parameters) and
the brightness of the background, $\bar{n}_{B}$. As shown in Fig.~\ref%
{qicrit}, QI's superiority prevails in a substantial region of $\Gamma _{%
\text{w}}$, $\Gamma _{o}$ values corresponding to Fig.~\ref{fig2} regions
where our source has the best efficiency in producing quantum entanglement
and, more generally, quantum correlations, per microwave photon emitted.
Such advantage is found as long as the average number of microwave photons
is sufficiently low.
\begin{figure}[h!]
\vspace{-0.7cm} \centering
\includegraphics[width=2.9in]{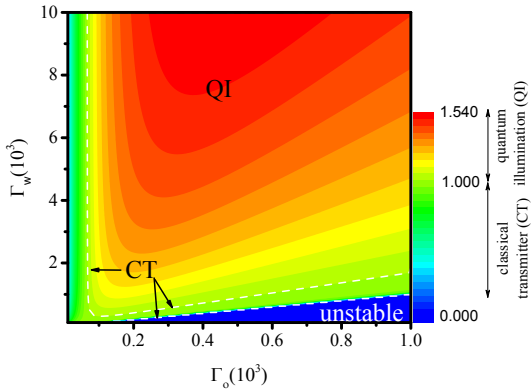}
\vspace{-0.3cm}
\caption{QI-advantage figure of merit, $\mathcal{F}$, versus $\Gamma _{%
\mathrm{w}}$ and $\Gamma _{o}$. For $\mathcal{F}>1$, the QI system
has lower error probability than any classical-state system, i.e.,
classical transmitter (CT), of the same transmitted energy. See
Fig. \protect\ref{fig2} for the other parameter values.}
\label{qicrit}
\end{figure}

\textit{Conclusion and Discussion}.--~We have shown that quantum
illumination can be performed in its more natural setting for
target detection, i.e., the microwave regime, by using a pair of
electro-opto-mechanical converters. Thanks to this converter, the
target region can be interrogated at a microwave frequency, while
the quantum-illumination joint measurement needed for target
detection is made at optical frequency, where the high-performance
photodetectors needed to obtain QI's performance advantage are
available.

An optimized EOM converter is able to generate strong quantum entanglement
and quantum correlations between its microwave and optical outputs. These
correlations can successfully be exploited for target detection, yielding
lower error probability than that of any classical-state microwave system of
the same transmitted energy. The QI advantage is especially evident when
detecting the faint returns from low-reflectivity objects embedded in the
bright thermal noise typical of room-temperature microwave environments.


Note that we assumed unit quantum efficiency for the optical part
of our quantum receiver. This is not far from current experimental
conditions: photon collection efficiencies from optical cavities
can be very high ($>74\%$ in Ref.~\cite{Rempe13}), loss at the
beam splitter can be extremely low, and photodetection can be
extremely efficient at optical wavelengths. Thus the main source
of loss may come from the optical storage of the idler mode, to be
preserved during the signal roundtrip time. This is not an issue
for short-range applications but, for long-range tasks, the idler
loss must remain below 3~dB, otherwise the QI advantage of the
phase-conjugating quantum receiver is lost~\cite{saikat}. While
using a good quantum memory (e.g., a rare-earth
doped-crystal~\cite{Zhong15}) would solve the problem, the
practical solution of storing the idler into an optical-fiber
delay line would restrict the maximum range of the quantum radar
to about $11.25$~km in free-space (assuming a fiber loss of
$0.2$~dB/km and fiber propagation speed equal to $2c/3$, where $c$
is vacuum light-speed).


Finally, extending our results to lower frequencies (below
1\thinspace GHz), our scheme could potentially be used for
non-invasive NMR spectroscopy in structural biology (structure of
proteins and nucleic acids) and in medical applications (magnetic
resonance imaging). Future implementations of quantum illumination
at the microwave regime could also be achieved by using other
quantum sources, for instance based on Josephson parametric
amplifiers, which are able to generate entangled microwave modes
of high quality~\cite{JPA1,JPA2,JPA3,JPA4}. These amplifiers might
become a very good choice once that suitable high-performance
microwave photo-detectors are made available.

\textit{Acknowledgments}.--~S.B. is grateful for support from the
Alexander von Humboldt foundation. D.V. is sponsored by the
European Commission (ITN-Marie Curie Project \lq cQOM\rq, Grant
No. 290161, and FET-Open Project \lq iQUOEMS\rq, Grant No.
323924). S.G. was supported by the US Office of Naval Research
contract number N00014-14-C-0002. J.H.S. appreciates sponsorship
by AFOSR and ONR. S.P. has been sponsored by a Leverhulme Trust
research fellowship (\lq qBIO\rq ) and EPSRC via \lq
qDATA\rq~(Grant no. EP/L011298/1) and \lq HIPERCOM\rq~(Grant No.
EP/J00796X/1).


\setcounter{section}{0} \setcounter{subsection}{0}
\renewcommand{\bibnumfmt}[1]{[S#1]} \renewcommand{\citenumfont}[1]{S#1}


\newpage

\widetext

\begin{center}
{\huge Supplemental Material}

\bigskip
\end{center}

\textbf{Contents}. In this Supplemental Material we provide
technical details on the following: (i) Hamiltonian of an
electro-opto-mechanical system, (ii) linearization of the
Hamiltonian, (iii) derivation of the quantum Langevin equations
with internal losses, {(iv) study of the microwave-optical
entanglement considering the entanglement metric, the logarithmic
negativity and the coherent information}, (v) general study of the
quantum correlations as quantified by quantum discord, and (vi)
analysis of the error probability for $M$ mode pairs.

\section{Hamiltonian of an electro-opto-mechanical system}

Our electro-opto-mechanical~(EOM) converter consists of a
mechanical resonator (MR) that is capacitively coupled on one side
to a driven
superconducting microwave cavity, and on the other side to a driven Fabry-P%
\'{e}rot optical cavity \cite{lenh1APP,bochmannAPP,bagciAPP}.
These cavities' driving fields are at radian frequencies $\omega
_{\mathrm{d},j}=\omega _{j}-\Delta _{0,j}$, where the $\Delta
_{0,j}$ are the detunings from their resonant frequencies $\omega
_{j}$, with $j=\text{w},o$ denoting the microwave and optical
cavities, respectively. We include intrinsic losses for these
cavities with rates $\kappa _{j}^{\mathrm{int}}$, and use $\kappa _{j}^{%
\mathrm{in}}$ to denote their input-port coupling rates. The
Hamiltonian of the coupled system in terms of annihilation and
creation operators has been studied in Ref. \cite{shabir2APP}, and
is given by
\begin{eqnarray}
\hat{H} &=&\hbar \omega _{M}\hat{b}^{\dagger }\hat{b}+\hbar \sum_{j=\mathrm{w%
},o}\omega _{j}\hat{a}_{j}^{\dagger }\hat{a}_{j}+\frac{\hbar g_{\mathrm{w}}}{%
2}(\hat{b}^{\dagger }+\hat{b})(\hat{a}_{\mathrm{w}}+\hat{a}_{\mathrm{w}%
}^{\dagger })^{2}+\hbar g_{o}(\hat{b}^{\dagger }+\hat{b})\hat{a}%
_{o}^{\dagger }\hat{a}_{o}  \notag  \label{ham1} \\
&&+\mathrm{i}\hbar E_{\mathrm{w}}(e^{\mathrm{i}\omega _{\mathrm{d,w}}t}-e^{-%
\mathrm{i}\omega _{\mathrm{d,w}}t})(\hat{a}_{\mathrm{w}}+\hat{a}_{\mathrm{w}%
}^{\dagger })+\mathrm{i}\hbar E_{o}(\hat{a}_{o}^{\dagger }e^{-\mathrm{i}%
\omega _{\mathrm{d},o}t}-\hat{a}_{o}e^{\mathrm{i}\omega
_{\mathrm{d},o}t}).
\end{eqnarray}
Here, $\hat{b}$ is the annihilation operator of the MR whose
resonant frequency is $\omega _{M}$, $\hat{a}_{j}$ is the
annihilation operator for cavity $j$ whose coupling rate to the MR
is $g_{j}$. The optical-driving
strength is $E_{o}=\sqrt{2P_{o}\kappa _{o}^{\mathrm{in}}/\hbar \omega _{%
\mathrm{d},o}}$, where $P_{o}$ is the laser power. The microwave
cavity is
driven by an electric potential with $e(t)=-\mathrm{i}\sqrt{2\hbar \omega _{%
\mathrm{w}L}L}\,E_{\mathrm{w}}(e^{\mathrm{i}\omega _{\mathrm{d,w}}t}-e^{-%
\mathrm{i}\omega _{\mathrm{d,w}}t})$ where $L$ is the inductance
in the microwave electric circuit and $E_{\mathrm{w}}$ describes
the amplitude of the microwave driving field~\cite{shabir2APP}. In
writing Eq.~(\ref{ham1}) we have ignored any additional degenerate
modes that the optical and microwave cavities might have. Ignoring
such modes is justified whenever scattering of photons from the
driven mode into other cavity modes can be neglected. That
condition prevails, for example, with short cavities whose free
spectral ranges greatly exceed the MR's resonant frequency, thus
ensuring that only one mode from each cavity interacts with the
MR, and that adjacent modes are not excited by the driving fields.

In the interaction picture with respect to $\hbar \omega _{\mathrm{d,w}}a_{%
\mathrm{w}}^{\dagger }a_{\mathrm{w}}+\hbar \omega _{\mathrm{d}%
,o}a_{o}^{\dagger }a_{o}$, and neglecting terms oscillating at
$\pm 2\omega _{\mathrm{d,w}}$, the system Hamiltonian reduces to
\begin{equation}
\hat{H}=\hbar \omega _{M}\hat{b}^{\dagger }\hat{b}+\hbar \sum_{j=\mathrm{w}%
,o}\Big[\Delta _{0,j}+g_{j}(\hat{b}^{\dagger }+\hat{b})\Big]\hat{a}%
_{j}^{\dagger }\hat{a}_{j}+\hat{H}_{\mathrm{dri}},  \label{ham2}
\end{equation}%
where the Hamiltonian associated with the driving fields is $\hat{H}_{%
\mathrm{dri}}=\mathrm{i}\hbar \sum_{j=\mathrm{w},o}E_{j}(\hat{a}%
_{j}^{\dagger }-\hat{a}_{j})$.

\section{Linearization of the Hamiltonian}

We can linearize Hamiltonian~(\ref{ham2}) by expanding the cavity
modes
around their steady-state field amplitudes, $\hat{c}_{j}=\hat{a}_{j}-\sqrt{%
N_{j}}$, where $N_{j}=|E_{j}|^{2}/(\kappa _{j}^{2}+\Delta
_{j}^{2})\gg 1$ is the mean number of intracavity photons induced
by the microwave or optical pumps~\cite{shabir2APP}, the $\kappa
_{j}=\kappa _{j}^{\mathrm{in}}+\kappa _{j}^{\mathrm{int}}$ are the
total cavity decay rates, and the $\Delta _{j}$ are the effective
cavity detunings. It is then convenient to move to the
interaction picture with respect to the free Hamiltonian, $\hbar \omega _{M}%
\hat{b}^{\dagger }\hat{b}+\hbar \sum_{j=\mathrm{w},o}\omega _{j}\hat{a}%
_{j}^{\dagger }\hat{a}_{j}$, where the linearized Hamiltonian
becomes
\begin{equation}
\hat{H}=\hbar \sum_{j=\mathrm{w},o}G_{j}(\hat{b}\mathrm{e}^{-\mathrm{i}%
\omega _{M}t}+\hat{b}^{\dagger }\mathrm{e}^{\mathrm{i}\omega _{M}t})(\hat{c}%
_{j}^{\dagger }\mathrm{e}^{\mathrm{i}\Delta _{j}t}+\hat{c}_{j}\mathrm{e}^{-%
\mathrm{i}\Delta _{j}t}),  \label{ham3}
\end{equation}%
where $G_{j}=g_{j}\sqrt{N_{j}}$. By setting the effective cavity
detunings so that $\Delta _{\mathrm{w}}=-\Delta _{{o}}=\omega
_{M}$ and neglecting the terms rotating at $\pm 2\omega _{M}$, the
above Hamiltonian reduces to
\begin{equation}
\hat{H}=\hbar G_{o}(\hat{c}_{o}\hat{b}+\hat{b}^{\dagger }\hat{c}%
_{o}^{\dagger })+\hbar G_{\mathrm{w}}(\hat{c}_{\mathrm{w}}\hat{b}^{\dagger }+%
\hat{b}\hat{c}_{\mathrm{w}}^{\dagger }),  \label{hameff}
\end{equation}%
as specified in the paper.

\section{Quantum Langevin equations with internal losses}

The full quantum treatment of the system can be given in terms of
the quantum Langevin equations in which we add to the Heisenberg
equations the quantum noise acting on the mechanical resonator
($\hat b_{\text{int}} $
with damping rate $\gamma_M$), as well as the cavities' input fluctuations ($%
\hat c_{j,\mathrm{\ in}}$, for $j=\mathrm{w},o$, with rates $\kappa_j^{%
\mathrm{in}}$), plus the intrinsic losses of the cavity modes ($\hat c_{j,%
\mathrm{\ int}}$, for $j=\mathrm{w},o$, with loss rates $\kappa_j^{\mathrm{%
int}}$). These noises have the correlation functions
\begin{subequations}
\begin{align}
\langle \hat c_{j,\mathrm{\ in}}(t) \hat c_{j,\mathrm{\ in}}
^{\dagger}(t^{\prime})\rangle & = \langle \hat c_{j,\mathrm{\ in}}
^{\dagger}(t) \hat c_{j,\mathrm{\ in}}(t^{\prime})\rangle
+\delta(t-t^{\prime})=(\bar{n}_j^T+1)\delta(t-t^{\prime}), \\
\langle \hat c_{j,\mathrm{\ int}}(t) \hat c_{j,\mathrm{\ int}}
^{\dagger}(t^{\prime})\rangle & = \langle \hat c_{j,\mathrm{\
int}} ^{\dagger}(t) \hat c_{j,\mathrm{\ int}}(t^{\prime})\rangle
+\delta(t-t^{\prime})=(\bar{n}_j^{\mathrm{int}}+1)\delta(t-t^{\prime}), \\
\langle \hat b_{\mathrm{\ int}}(t) \hat b_{\mathrm{\ int}}
^{\dagger}(t^{\prime})\rangle & = \langle \hat b_{\mathrm{\ int}%
}^{\dagger}(t) \hat b_{\mathrm{\ int}}(t^{\prime})\rangle
+\delta(t-t^{\prime})=(\bar{n}_b^T+1)\delta(t-t^{\prime}),
\end{align}
where $\bar{n}_j^{\mathrm{\ int}}$, $\bar{n}_j^T$, and
$\bar{n}_b^T$ are the Planck-law thermal occupancies of each bath.
The resulting Langevin equations for the cavity modes and MR are
\end{subequations}
\begin{subequations}
\begin{align}  \label{qles2a}
\hat{\dot{c}}_{\mathrm{w}}&=-\kappa_{\mathrm{w}} \hat c_{\mathrm{w}}-\mathrm{%
i}G_{\mathrm{w}} \hat b+\sqrt{2\kappa_{\mathrm{w}}^{\mathrm{in}}}\hat c_{%
\mathrm{w,in}}+\sqrt{2\kappa_{\mathrm{w}}^{\mathrm{int}}}\hat c_{\mathrm{%
w,int}}, \\
\hat{\dot{c}}_{{o}}&=-\kappa_{{o}} \hat c_{{o}}-\mathrm{i}G_{{o}}
\hat
b^{\dagger}+\sqrt{2\kappa_{{o}}^{\mathrm{in}}}\hat c_{o,\mathrm{in}}+\sqrt{%
2\kappa_{{o}}^{\mathrm{int}}}\hat c_{o,\mathrm{int}}, \\
\hat{\dot{b}}&=-\gamma_M \hat b-\mathrm{i}G_{{o}} \hat c_{{o}}^{\dagger}-%
\mathrm{i}G_{\mathrm{w}} \hat c_{\mathrm{w}}+\sqrt{2\gamma_M}\hat b_{\mathrm{%
\ int}}.  \label{qles2c}
\end{align}
We can solve the above equations in the Fourier domain to obtain
the microwave and optical cavities' variables. By substituting the
solutions of Eqs.~(\ref{qles2a})--(\ref{qles2c}) into the
corresponding input-output
formula for the cavities' variables, i.e., $\hat d_j\equiv \hat c_{j,\text{%
out}}= \sqrt{2\kappa_j^{\mathrm{in}}}\hat c_j-\hat
c_{j,\mathrm{in}}$, we obtain
\end{subequations}
\begin{subequations}
\begin{align}  \label{qlessimplifya}
\hat d_{\mathrm{w}}&= A_{\mathrm{w}}(\omega)\hat c_{\mathrm{w,in}%
}-B(\omega)\hat c_{o,\mathrm{in}}^{\dagger}-C_{\mathrm{w}}(\omega) \hat b_{%
\mathrm{int}}-D_{\mathrm{w}}(\omega) \hat c_{o,\mathrm{int}}^{\dagger}-E_{%
\mathrm{w}}(\omega) \hat c_{\mathrm{w,int}}, \\
\hat d_{{o}}&= A_{{o}}(\omega)\hat c_{o,\mathrm{in}}+B^*(\omega)\hat c_{%
\mathrm{w,in}}^{\dagger}-C_{{o}}(\omega) \hat b_{\mathrm{int}}^{\dagger}+D_{{%
o}}(\omega) \hat
c_{\mathrm{w},\mathrm{int}}^{\dagger}+E_{{o}}(\omega) \hat
c_{o,\mathrm{int}},  \label{qlessimplifyc}
\end{align}
where
\end{subequations}
\begin{subequations}
\begin{align}  \label{coeffa}
A_{\mathrm{w}}(\omega)&=\frac{[\tilde{\omega}_{\mathrm{w}}-2\kappa_{\mathrm{w%
}}^{\mathrm{in}}/\kappa_\mathrm{w}][\Gamma_{{o}}-\tilde{\omega}_{{o}}\tilde{%
\omega}_b]-\Gamma_{\mathrm{w}}\tilde{\omega}_{{o}}}{\tilde{\omega}_{\mathrm{w%
}}[\tilde{\omega}_{{o}}\tilde{\omega}_b-\Gamma_{{o}}]+\Gamma_{\mathrm{w}}%
\tilde{\omega}_{{o}}}, \\
A_{{o}}(\omega)&=\frac{-[\tilde{\omega}_{{o}}^*-2\kappa_{{o}}^{\mathrm{in}%
}/\kappa_{o}][\Gamma_{\mathrm{w}}+\tilde{\omega}_{\mathrm{w}}^*\tilde{\omega}%
_b^*]+\Gamma_{{o}}\tilde{\omega}^*_{\mathrm{w}}}{\tilde{\omega}^*_{\mathrm{w}%
}[\tilde{\omega}^*_{{o}}\tilde{\omega}^*_b-\Gamma_{{o}}]+\Gamma_{\mathrm{w}}%
\tilde{\omega}^*_{{o}}}, \\
B(\omega)&=2\sqrt{\frac{\kappa_{{o}}^{\mathrm{in}}}{\kappa_{{o}}}}\sqrt{%
\frac{\kappa_{\mathrm{w}}^{\mathrm{in}}}{\kappa_{\mathrm{w}}}}\frac{\sqrt{%
\Gamma_{\mathrm{w}}\Gamma_{{o}}}}{\tilde{\omega}_{\mathrm{w}}[\tilde{\omega}%
_{{o}}\tilde{\omega}_b-\Gamma_{{o}}]+\Gamma_{\mathrm{w}}\tilde{\omega}_{{o}}}%
, \\
C_{\mathrm{w}}(\omega)&=\sqrt{\frac{\kappa_{\mathrm{w}}^{\mathrm{in}}}{%
\kappa_{\mathrm{w}}}}\frac{2\mathrm{i}\sqrt{\Gamma_{\mathrm{w}}}\tilde{\omega%
}_{{o}}}{\tilde{\omega}_{\mathrm{w}}[\tilde{\omega}_{{o}}\tilde{\omega}%
_b-\Gamma_{{o}}]+\Gamma_{\mathrm{w}}\tilde{\omega}_{{o}}}, \\
C_{{o}}(\omega)&=\sqrt{\frac{\kappa_{{o}}^{\mathrm{in}}}{\kappa_{{o}}}}\frac{%
2\mathrm{i}\sqrt{\Gamma_{{o}}}\tilde{\omega}^*_{\mathrm{w}}}{\tilde{\omega}%
^*_{\mathrm{w}}[\tilde{\omega}^*_{{o}}\tilde{\omega}^*_b-\Gamma_{{o}%
}]+\Gamma_{\mathrm{w}}\tilde{\omega}^*_{{o}}}, \\
D_{\mathrm{w}}(\omega)&=2\sqrt{\frac{\kappa_{{o}}^{\mathrm{int}}}{\kappa_{{o}%
}}}\sqrt{\frac{\kappa_{\mathrm{w}}^{\mathrm{in}}}{\kappa_{\mathrm{w}}}}\frac{%
\sqrt{\Gamma_{{o}}\Gamma_{\mathrm{w}}}}{\tilde{\omega}_{\mathrm{w}}[\tilde{%
\omega}_{{o}}\tilde{\omega}_b-\Gamma_{{o}}]+\Gamma_{\mathrm{w}}\tilde{\omega}%
_{{o}}}, \\
D_{{o}}(\omega)&=2\sqrt{\frac{\kappa_{{o}}^{\mathrm{in}}}{\kappa_{{o}}}}%
\sqrt{\frac{\kappa_{\mathrm{w}}^{\mathrm{int}}}{\kappa_{\mathrm{w}}}}\frac{%
\sqrt{\Gamma_{{o}}\Gamma_{\mathrm{w}}}}{\tilde{\omega}^*_{\mathrm{w}}[\tilde{%
\omega}^*_{{o}}\tilde{\omega}^*_b-\Gamma_{{o}}]+\Gamma_{\mathrm{w}}\tilde{%
\omega}^*_{{o}}}, \\
E_{\mathrm{w}}(\omega)&=2\sqrt{\frac{\kappa_{\mathrm{w}}^{\mathrm{int}}}{%
\kappa_{\mathrm{w}}}}\sqrt{\frac{\kappa_{\mathrm{w}}^{\mathrm{in}}}{\kappa_{%
\mathrm{w}}}}\frac{\Gamma_{{o}}-\tilde{\omega}_{{o}}\tilde{\omega}_b}{\tilde{%
\omega}_{\mathrm{w}}[\tilde{\omega}_{{o}}\tilde{\omega}_b-\Gamma_{{o}%
}]+\Gamma_{\mathrm{w}}\tilde{\omega}_{{o}}}, \\
E_{{o}}(\omega)&=2\sqrt{\frac{\kappa_{{o}}^{\mathrm{int}}}{\kappa_{{o}}}}%
\sqrt{\frac{\kappa_{{o}}^{\mathrm{in}}}{\kappa_{{o}}}}\frac{\Gamma_{\mathrm{w%
}}+\tilde{\omega}^*_{\mathrm{w}}\tilde{\omega}^*_b}{\tilde{\omega}^*_{%
\mathrm{w}}[\tilde{\omega}^*_{{o}}\tilde{\omega}^*_b-\Gamma_{{o}}]+\Gamma_{%
\mathrm{w}}\tilde{\omega}^*_{{o}}},  \label{coeffi}
\end{align}
with $\tilde{\omega}_j=1-\mathrm{i}\omega/\kappa_j$, $\tilde{\omega}_b=1-%
\mathrm{i}\omega/\gamma_M$, and $\Gamma_j =
G^2_j/\kappa_j\gamma_M$. The
coefficients (\ref{coeffa})--(\ref{coeffi}) become much simpler at $%
\omega\simeq 0 $, which corresponds to take a narrow frequency
band around each cavity resonance, viz.,
\end{subequations}
\begin{subequations}
\begin{align}  \label{coef2}
A_{\mathrm{w}}&=\frac{[1-2\kappa_{\mathrm{w}}^{\mathrm{in}}/\kappa_\mathrm{w}%
][\Gamma_{{o}}-1]-\Gamma_{\mathrm{w}}}{1-\Gamma_{{o}}+\Gamma_{\mathrm{w}}},
\\
A_{{o}}&=\frac{-[1-2\kappa_{{o}}^{\mathrm{in}}/\kappa_{o}][\Gamma_{\mathrm{w}%
}+1]+\Gamma_{{o}}}{1-\Gamma_{{o}}+\Gamma_{\mathrm{w}}}, \\
B&=2\sqrt{\frac{\kappa_{{o}}^{\mathrm{in}}}{\kappa_{{o}}}}\sqrt{\frac{%
\kappa_{\mathrm{w}}^{\mathrm{in}}}{\kappa_{\mathrm{w}}}}\frac{\sqrt{\Gamma_{%
\mathrm{w}}\Gamma_{{o}}}}{1-\Gamma_{{o}}+\Gamma_{\mathrm{w}}}, \\
C_{\mathrm{w}}&=\sqrt{\frac{\kappa_{\mathrm{w}}^{\mathrm{in}}}{\kappa_{%
\mathrm{w}}}}\frac{2\mathrm{i}\sqrt{\Gamma_{\mathrm{w}}}}{1-\Gamma_{{o}%
}+\Gamma_{\mathrm{w}}}, \\
C_{{o}}&=\sqrt{\frac{\kappa_{{o}}^{\mathrm{in}}}{\kappa_{{o}}}}\frac{2%
\mathrm{i}\sqrt{\Gamma_{{o}}}}{1-\Gamma_{{o}}+\Gamma_{\mathrm{w}}}, \\
D_{\mathrm{w}}&=2\sqrt{\frac{\kappa_{{o}}^{\mathrm{int}}}{\kappa_{{o}}}}%
\sqrt{\frac{\kappa_{\mathrm{w}}^{\mathrm{in}}}{\kappa_{\mathrm{w}}}}\frac{%
\sqrt{\Gamma_{{o}}\Gamma_{\mathrm{w}}}}{1-\Gamma_{{o}}+\Gamma_{\mathrm{w}}},
\\
D_{{o}}&=2\sqrt{\frac{\kappa_{{o}}^{\mathrm{in}}}{\kappa_{{o}}}}\sqrt{\frac{%
\kappa_{\mathrm{w}}^{\mathrm{int}}}{\kappa_{\mathrm{w}}}}\frac{\sqrt{\Gamma_{%
{o}}\Gamma_{\mathrm{w}}}}{1-\Gamma_{{o}}+\Gamma_{\mathrm{w}}}, \\
E_{\mathrm{w}}&=2\sqrt{\frac{\kappa_{\mathrm{w}}^{\mathrm{int}}}{\kappa_{%
\mathrm{w}}}}\sqrt{\frac{\kappa_{\mathrm{w}}^{\mathrm{in}}}{\kappa_{\mathrm{w%
}}}}\frac{\Gamma_{{o}}-1}{1-\Gamma_{{o}}+\Gamma_{\mathrm{w}}}, \\
E_{{o}}&=2\sqrt{\frac{\kappa_{{o}}^{\mathrm{int}}}{\kappa_{{o}}}}\sqrt{\frac{%
\kappa_{{o}}^{\mathrm{in}}}{\kappa_{{o}}}}\frac{\Gamma_{\mathrm{w}}+1}{%
1-\Gamma_{{o}}+\Gamma_{\mathrm{w}}}.
\end{align}
Furthermore, when the internal losses are negligible, i.e., $\kappa^{\mathrm{%
int}}_j/\kappa_j^{\mathrm{in}} \ll 1$, then we get $D_j=E_j\simeq
0 $, and Eqs.~(\ref{qlessimplifya})--(\ref{qlessimplifyc}) reduce
to the simple forms presented in the paper,
\end{subequations}
\begin{subequations}
\begin{align}
\hat{d}_{\mathrm{w}} & =A_{\mathrm{w}}\hat{c}_{\mathrm{w,in}}-B\hat {c}_{o,%
\mathrm{in}}^{\dagger}-C_{\mathrm{w}}\hat{b}_{\mathrm{int}},  \label{qle1} \\
\hat{d}_{\mathrm{o}} & =B\hat{c}_{{\mathrm{w,in}},}^{\dagger}+A_{\mathrm{o}}%
\hat{c}_{o,\mathrm{in}}-C_{\mathrm{o}}\hat{b}_{\mathrm{int}}^{\dagger},
\label{simpleb}
\end{align}
with coefficients given by
\end{subequations}
\begin{subequations}
\begin{align}  \label{papercoeffa}
A_{\mathrm{w}} &=\frac{1-(\Gamma_{\mathrm{w}}+\Gamma_{o})}{1+\Gamma_{\mathrm{%
w}}-\Gamma_{o}}, \\
A_{o} &=\frac{1+(\Gamma_{\mathrm{w}}+\Gamma_{o})}{1+\Gamma_{\mathrm{w}%
}-\Gamma_{o}}, \\
B &=\frac{2\sqrt{\Gamma_{\mathrm{w}}\Gamma_{o}}}{1+\Gamma_{\mathrm{w}%
}-\Gamma_{o}}, \\
C_{o} &=\frac {2\mathrm{i}\sqrt{\Gamma_{o}}}{1+\Gamma_{\mathrm{w}}-\Gamma_{o}%
}, \\
C_{\text{w}} &=\frac {2\mathrm{i}\sqrt{\Gamma_{\text{w}}}}{1+\Gamma_{\mathrm{%
w}}-\Gamma_{o}}.  \label{papercoeffc}
\end{align}
These input-output relations preserve the bosonic commutation
relations,
i.e., when the operators on the right in Eqs.~(\ref{qle1}) and (\ref{simpleb}%
) satisfy those commutation relations, we get $[\hat d_i,\hat
d_j^\dagger]=\delta_{i,j}$ and $[\hat d_i,\hat d_j]=[\hat
d_i^\dagger,\hat d_j^\dagger]=0$, for $i,j\in \mathrm{w},o$.

\section{Microwave-optical entanglement}

\subsection{Entanglement Metric}

Equations~(\ref{qle1}) and (\ref{simpleb}) are driven by a
collection of
independent, thermal-state inputs, $\hat{c}_{\text{w,in}}$, $\hat{c}_{o,%
\text{in}}$, and $\hat{b}_{\text{int}}$ and their adjoints.
Consequently, those equations' outputs, $\hat{d}_{\text{w}}$ and
$\hat{d}_{o}$, are in a zero-mean, jointly-Gaussian state that is
completely characterized by its non-zero second moments,
\end{subequations}
\begin{subequations}
\begin{align}
\bar{n}_{\text{w}}& \equiv \langle \hat{d}_{\text{w}}^{\dagger }\hat{d}_{%
\text{w}}\rangle =|A_{\text{w}}|^{2}\bar{n}_{\text{w}}^{T}+|B|^{2}(\bar{n}%
_{o}^{T}+1)+|C_{\text{w}}|^{2}\bar{n}_{b}^{T},  \label{dwdw} \\
\bar{n}_{o}& \equiv \langle \hat{d}_{o}^{\dagger
}\hat{d}_{o}\rangle
=|B|^{2}(\bar{n}_{\text{w}}^{T}+1)+|A_{o}|^{2}\bar{n}_{o}^{T}+|C_{o}|^{2}(%
\bar{n}_{b}^{T}+1),  \label{dodo} \\
\langle \hat{d}_{\text{w}}\hat{d}_{o}\rangle & =A_{\text{w}}B(\bar{n}_{\text{%
w}}^{T}+1)-BA_{o}\bar{n}_{o}^{T}+C_{\text{w}}C_{o}(\bar{n}_{b}^{T}+1).
\label{dwdo}
\end{align}

That joint state will be \emph{classical}, i.e., have a proper $P$%
-representation, if and only if the phase-sensitive cross correlation, $%
|\langle \hat{d}_{\text{w}}\hat{d}_{o}\rangle |$, satisfies the
classical bound~\cite{footnote1}
\end{subequations}
\begin{equation}
|\langle \hat{d}_{\text{w}}\hat{d}_{o}\rangle |\leq \sqrt{\bar{n}_{\text{w}}%
\bar{n}_{o}}.
\end{equation}%
When $|\langle \hat{d}_{\text{w}}\hat{d}_{o}\rangle |$ violates
this bound, the microwave (w) and optical ($o$) modes are
entangled. Note that, regardless of whether the \emph{classical}
bound is violated, the absence of
phase-sensitive second moments in the joint state, viz., $\langle \hat{d}_{%
\text{w}}^{2}\rangle =\langle \hat{d}_{o}^{2}\rangle =0$, implies
that the reduced density operators for these individual modes are
thermal states. The
average photon numbers for the microwave and optical modes also put a \emph{%
quantum} upper limit on $|\langle
\hat{d}_{\text{w}}\hat{d}_{o}\rangle |$, given by~\cite{footnote2}
\begin{equation}
|\langle \hat{d}_{\text{w}}\hat{d}_{o}\rangle |\leq \sqrt{\max (\bar{n}_{%
\text{w}},\bar{n}_{o})[1+\min (\bar{n}_{\text{w}},\bar{n}_{o})]},
\label{qbound}
\end{equation}%
which can be saturated when $\bar{n}_{\text{w}}=\bar{n}_{o}$, but not when $%
\bar{n}_{\text{w}}\neq \bar{n}_{o}$.

Figure~2 in the main paper plots---versus $\Gamma _{\text{w}}$ and
$\Gamma _{o}$, and assuming experimentally-accessible parameter
values---the entanglement metric
\begin{equation}
\mathcal{E}\equiv \frac{|\langle \hat{d}_{\text{w}}\hat{d}_{o}\rangle |}{%
\sqrt{\bar{n}_{\text{w}}\bar{n}_{o}}},
\end{equation}%
which exceeds unity if and only if the microwave and optical modes
are entangled. The entanglement metric provides a very simple
entanglement criterion and, as we can see from the figure in the
main paper, our
microwave-optical source is proven to have wide regions of parameters where $%
\mathcal{E}>1$.

\subsection{Logarithmic Negativity and Coherent Information}

Here we quantify the amount of entanglement generated by our
microwave-optical source using standard measures in quantum
information theory. In particular, we consider the
log-negativity~\cite{eisAPP} and the coherent
information~\cite{cohinfo}, which are, respectively, an upper and
a lower bound to the number of distillable entanglement bits
(ebits) generated by the source. We normalize these measures by
the mean number of microwave signal photons $\bar{n}_{\text{w}}$
sent to the target. The rationale behind this normalization is
physical and strictly connected with the specific model
considered.

In fact, quantum illumination (QI) is a energy-constrained
protocol, where quantum and classical sources are compared by
fixing the mean number of photons in the signal
mode~\cite{pirandolaAPP}, which corresponds to the microwave mode
in the present case. Thus, it is intuitively expected that, at
fixed $\bar{n}_{\text{w}}$, the advantage of quantum illumination
increases by increasing the amount of quantum entanglement (and
more generally, quantum correlations) present in the source.

For this reason, the quality of the source can be estimated in two
ways: (1) At fixed $\bar{n}_{\text{w}}$, we optimize the
entanglement of the quantum source, or (2) we maximize the
entanglement of the quantum source normalized by
$\bar{n}_{\text{w}}$, i.e., the average entanglement carried by
each
microwave photon. The two options are clearly connected since, at fixed $%
\bar{n}_{\text{w}}$, the most entangled source is the one emitting
more ebits per microwave photon. However, the second option is
more flexible to use, since it can be applied to the cases where
$\bar{n}_{\text{w}}$ changes in terms of the various parameters of
the problem, as it happens here on the cooperativity plane
($\Gamma _{o}$,$\Gamma _{\text{w}}$).

\subsubsection{Logarithmic Negativity}

In order to compute the log-negativity, we first determine the
covariance matrix~(CM) of our system in the frequency domain,
which can be expressed as
\begin{equation}
\delta (\omega +\omega ^{\prime })V_{ij}(\omega
)=\frac{1}{2}\langle u_{i}(\omega )u_{j}(\omega ^{\prime
})+u_{j}(\omega ^{\prime })u_{i}(\omega )\rangle ,  \label{cor1}
\end{equation}%
where
\begin{equation}
\mathbf{u}(\omega )=[X_{\mathrm{w}}(\omega ),Y_{\mathrm{w}}(\omega
),X_{o}(\omega ),Y_{o}(\omega )]^{T},
\end{equation}%
and $X_{j}=(d_{j}+d_{j}^{\dagger })/\sqrt{2}$,
$Y_{j}=(d_{j}-d_{j}^{\dagger })/\mathrm{i}\sqrt{2}$ with
$j=o,\mathrm{w}$. Note that the vacuum noise has variance $1/2$ in
these quadratures.

Now, by using Eqs.~(\ref{qle1}), (\ref{simpleb}) and~(\ref{cor1}),
we obtain the CM for the quadratures of the{\ microwave and
optical cavities' outputs, which is given by the normal form
\begin{equation}
\mathbf{V}(\omega )=\left(
\begin{array}{cccc}
V_{11} & 0 & V_{13} & 0 \\
0 & V_{11} & 0 & -V_{13} \\
V_{13} & 0 & V_{33} & 0 \\
0 & -V_{13} & 0 & V_{33}%
\end{array}
\right) ,  \label{driftA}
\end{equation}
where we explicitly have
\begin{subequations}
\begin{align}
V_{11}& =\frac{\langle X_{\mathrm{w}}(\omega
)X_{\mathrm{w}}(\omega ^{\prime
})\rangle }{\delta (\omega +\omega ^{\prime })}=\bar{n}_{\text{w}}+1/2, \\
V_{33}& =\frac{\langle X_{o}(\omega )X_{o}(\omega ^{\prime })\rangle }{%
\delta (\omega +\omega ^{\prime })}=\bar{n}_{o}+1/2, \\
V_{13}& =\frac{\langle X_{\mathrm{w}}(\omega )X_{o}(\omega
^{\prime })+X_{o}(\omega ^{\prime })X_{\mathrm{w}}(\omega )\rangle
}{2\delta (\omega +\omega ^{\prime })}=\langle
\hat{d}_{\mathrm{w}}\hat{d}_{o}\rangle ,
\end{align}
and we have used the fact that $\langle
\hat{d}_{\mathrm{w}}\hat{d}_{o}\rangle $ is real valued. Note that
Eq.~(\ref{driftA}) is the typical CM of a two-mode squeezed
thermal state~\cite{DiscordAPP}.

The log-negativity $E_{N}$ is given by~\cite{eisAPP}
\end{subequations}
\begin{equation}
E_{N}=\mathrm{max}[0,-\mathrm{\log }(2\zeta ^{-})],  \label{loga}
\end{equation}%
where $\zeta ^{-}$ is the smallest partially-transposed symplectic
eigenvalue of $\mathbf{V}(\omega )$, given by~\cite{RMP}%
\begin{equation}
\zeta ^{-}=2^{-1/2}\left( V_{11}^{2}+V_{33}^{2}+2V_{13}^{2}-\sqrt{%
(V_{11}^{2}-V_{33}^{2})^{2}+4V_{13}^{2}(V_{11}+V_{33})^{2}}\right)
^{1/2}.
\end{equation}%
In Fig.~2 of the main paper, we have plotted the normalized log-negativity $%
E_{N}/\bar{n}_{\text{w}}$ versus the cooperativities, $\Gamma
_{\text{w}}$ and $\Gamma _{o}$. From panel (b) of that figure, we
can see the presence of a wide region where the quality of the
source is very good in terms of ebits of log-negativity per
microwave photon emitted.

It is easy to show that $E_{N}=0$ for $\mathcal{E}\leq 1$.
Moreover, for the
case of interest for QI---in which $\bar{n}_{j}\equiv \langle \hat{d}%
_{j}^{\dagger }\hat{d}_{j}\rangle \leq 1$ for
$j=\mathrm{w},o$---we find that $2\zeta ^{-}$ decreases
monotonically, at fixed $\bar{n}_{\text{w}}$ and $\bar{n}_{o}$, as
$|\langle \hat{d}_{\mathrm{w}}\hat{d}_{o}\rangle |$ increases from
zero to its quantum upper bound from Eq.~(\ref{qbound}). Thus,
given $\bar{n}_{j}\leq 1$ for $j=\mathrm{w},o$, the logarithmic
negativity $E_{N}$ for our source can be related with the
entanglement metric $\mathcal{E}$.

\subsubsection{Coherent Information}

Here we compute the coherent information associated with our
microwave-optical source. This is given by~\cite{cohinfo}%
\begin{equation}
I(o\rangle \text{w})=S(\text{w})-S(o,\text{w}),
\label{cohinfoAPP}
\end{equation}%
where $S($w$)$ is the von Neumann entropy~\cite{RMP} of the
microwave mode, while $S(o,$w$)$ is the joint von Neumann entropy
of the optical and microwave modes. As mentioned before, this
provides a lower bound to the number of ebits which are
distillable from the source. In fact, according to the hashing
inequality~\cite{hashing}, $I(o\rangle $w$)$ gives the
distillation rate (ebits per use of the source) which is
achievable by distillation protocols based on one-way classical
communication (here directed from the optical to the microwave
part).

For a Gaussian state, like our source, we can simply express the
entropic
terms in Eq.~(\ref{cohinfoAPP}) in terms of the symplectic eigenvalue $\nu _{%
\text{w}}$ of the reduced CM associated with the microwave mode
and the
symplectic spectrum $\{\nu _{-},\nu _{+}\}$\ of the global CM $\mathbf{V}%
(\omega )$ associated with the microwave and optical modes. These
eigenvalues are given by $\nu _{\text{w}}=V_{11}$ and~\cite{RMP}%
\begin{equation}
\nu _{\pm }=2^{-1/2}\left( V_{11}^{2}+V_{33}^{2}-2V_{13}^{2}\pm \sqrt{%
(V_{11}^{2}-V_{33}^{2})^{2}-4V_{13}^{2}(V_{11}-V_{33})^{2}}\right)
^{1/2}.
\end{equation}%
Thus, we have~\cite{RMP}%
\begin{equation}
I(o\rangle \text{w})=h(V_{11})-h(\nu _{-})-h(\nu _{+}),
\end{equation}%
where~\cite{hx}
\begin{equation}
h(x)\equiv \left( x+\frac{1}{2}\right) \mathrm{log}_{2}\left( x+\frac{1}{2}%
\right) -\left( x-\frac{1}{2}\right) \mathrm{log}_{2}\left( x-\frac{1}{2}%
\right) .
\end{equation}

In Fig.~2 of the main paper, we have plotted the normalized
coherent information $I(o\rangle $w$)/\bar{n}_{\text{w}}$ versus
the cooperativities, $\Gamma _{\text{w}}$ and $\Gamma _{o}$. From
panel (c) of that figure, we can see the presence of a wide region
where the quality of the source is good in terms of qubits of
coherent information per microwave photon emitted (lower bound to
the number of ebits per microwave photon). Finally, note
that the computation of the normalized \textit{reverse} coherent information~%
\cite{rev1,rev2} $I(o\langle $w$)/\bar{n}_{\text{w}}$ (where the
direction of the classical communication is inverted from the
microwave to the optical part) leads to completely similar
results, as shown in Fig.~\ref{RevCoh}.
\begin{figure}[th]
\centering
\includegraphics[width=3.5in]{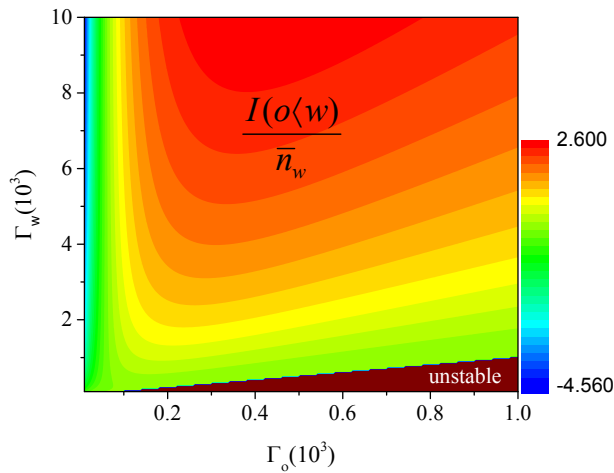}
\caption{Normalized reverse coherent information $I(o\langle $w$)/\bar{n}_{%
\text{w}}$ (qubits per microwave photon) plotted on the cooperativity plane (%
$\Gamma _{o},\Gamma _{\text{w}}$). See Fig. 2 of the main paper
for the other parameter values.} \label{RevCoh}
\end{figure}

\section{Microwave-optical quantum correlations beyond entanglement}

Our microwave-optical source generates a Gaussian state which is
mixed, as
one can easily check from the numerical values of its von Neumann entropy $%
S(o,$w$)$. It is therefore important to describe its quality also
in terms of general quantum correlations beyond quantum
entanglement. Thus we compute here the quantum
discord~\cite{DiscordRMP} of the source $D($w$|o)$, capturing the
basic quantum correlations which are carried by the microwave mode
sent to target. Such quantity is normalized by
$\bar{n}_{\text{w}}$, for the same reasons we have previously
explained for quantum entanglement: The best sources for our QI
problem are expected to be those maximizing the amount of quantum
correlations per microwave photon emitted.

Since our source emits a mixed Gaussian state which is a two-mode
squeezed thermal state, we can compute its (unrestricted) quantum
discord using the formulas of Ref.~\cite{DiscordAPP}. In
particular, the CM in Eq.~(\ref{driftA}) can be expressed
as~\cite{DiscordAPP}
\begin{equation}
\mathbf{V}(\omega )=\left(
\begin{array}{cc}
(\tau b+\eta )\mathbf{I} & \sqrt{\tau (b^{2}-1)}\mathbf{Z} \\
\sqrt{\tau (b^{2}-1)}\mathbf{Z} & b\mathbf{I}%
\end{array}%
\right) ,~~%
\begin{array}{c}
\mathbf{I}\equiv \mathrm{diag}(1,1),~~ \\
\mathbf{Z}\equiv \mathrm{diag}(1,-1),%
\end{array}%
\end{equation}%
where%
\begin{equation}
b=V_{33},~~\tau =\frac{V_{13}^{2}}{V_{33}^{2}-1},~~\eta =V_{11}-\frac{%
V_{33}V_{13}^{2}}{V_{33}^{2}-1}.
\end{equation}%
Thus, we may write~\cite{DiscordAPP}
\begin{eqnarray}
D(\text{w}|o) &=&h(b)-h(\nu _{-})-h(\nu _{+})+h(\tau +\eta ) \\
&=&h(V_{33})-h(\nu _{-})-h(\nu _{+})+h\left[ V_{11}+\frac{%
V_{13}^{2}(1-V_{33})}{V_{33}^{2}-1}\right] ,
\end{eqnarray}%
where $\nu _{-}$ and $\nu _{+}$ are the symplectic eigenvalues of $\mathbf{V}%
(\omega )$.

In Fig.~2 of the main paper, we have plotted the normalized quantum discord $%
D($w$|o)/\bar{n}_{\text{w}}$ versus the cooperativities, $\Gamma
_{\text{w}}$ and $\Gamma _{o}$. From panel (d) of that figure, we
can see the presence of a wide region where the quality of the
source is very good in terms of discordant bits per microwave
photon emitted. Furthermore, the similarity between this
normalized measure and the final QI avantage (figure 4 of the main
paper) is remarkable.

Finally, note that the computation of the normalized version of
the other quantum discord~\cite{DiscordRMP}
$D(o|$w$)/\bar{n}_{\text{w}}$ leads to similar results, as shown
below in Fig.~\ref{OtherDiscord} of this Supplemental Material.
\begin{figure}[th]
\centering
\includegraphics[width=3.5in]{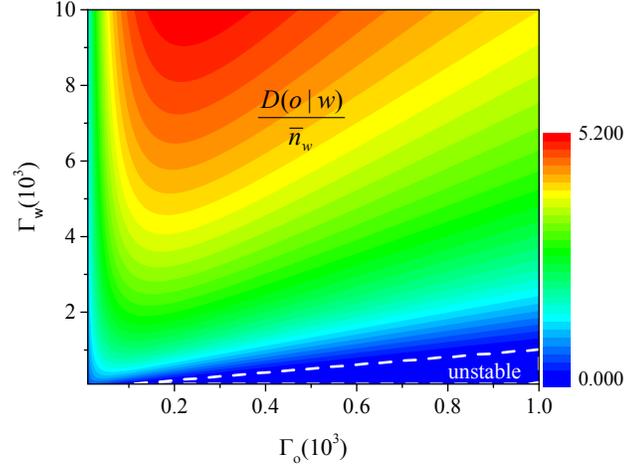}
\caption{Normalized quantum discord $D(o|$w$)/\bar{n}_{\text{w}}$
(discordant bits per microwave photon) plotted on the cooperativity plane ($%
\Gamma _{o},\Gamma _{\text{w}}$). See Fig. 2 of the main paper for
the other parameter values.} \label{OtherDiscord}
\end{figure}

\section{Error probability for $M$ mode pairs}

Our transmitter's EOM converter is governed by Eqs.~(\ref{qle1}) and (\ref%
{simpleb}) in which the coefficients appearing therein are as
specified in Eqs.~(\ref{papercoeffa})--(\ref{papercoeffc}). These
same coefficients appear in the input-output relation
\begin{equation}
\hat{d}_{\eta ,\mathrm{o}}=B\hat{c}_{R}^{\dagger }+A_{\mathrm{o}}\hat{c}_{%
\mathrm{o,in}}^{\prime
}-C_{\mathrm{o}}\hat{b}_{\text{int}}^{^{\prime }\dagger },
\label{phaseconjug}
\end{equation}%
for the QI receiver's EOM converter. Equations~(\ref{qle1}) and (\ref%
{simpleb}) apply to one of the $M=t_{m}W_{m}\gg 1$ independent,
identically-distributed (iid) mode pairs produced by the
bandwidth-$W_{m}$, continuous-wave transmitter during the
$t_{m}$-sec-duration measurement interval used to discriminate
target absence from target presence. Under either hypothesis
(target absent or present), the QI receiver takes as its inputs
$M$ iid mode pairs, $\{\hat{c}_{R}^{(k)},\hat{d}_{o}^{(k)}:1\leq
k\leq M\}$ that are in a zero-mean, jointly-Gaussian state
characterized by the conditional (given target absence or
presence) second moments: $\langle
\hat{c}_{R}^{(k)\dagger }\hat{c}_{R}^{(k)}\rangle _{H_{j}}$, $\langle \hat{d}%
_{o}^{(k)\dagger }\hat{d}_{o}^{(k)}\rangle $, and $\langle \hat{c}_{R}^{(k)}%
\hat{d}_{o}^{(k)}\rangle _{H_{j}}$. The receiver's EOM converter
transforms
the returned microwave modes, $\hat{c}_{R}^{(k)}$, into optical modes, $\hat{%
d}_{\eta ,o}^{(k)}$, such that the $\{\hat{d}_{\eta ,o}^{(k)},\hat{d}%
_{o}^{(k)}:1\leq k\leq M\}$ are iid mode pairs which, under either
hypothesis, are in a zero-mean jointly-Gaussian state
characterized by the second moments $\langle \hat{d}_{\eta
,o}^{(k)\dagger }\hat{d}_{\eta
,o}^{(k)}\rangle _{H_{j}}$, $\langle \hat{d}_{o}^{(k)\dagger }\hat{d}%
_{o}^{(k)}\rangle $, and $\langle \hat{d}_{\eta ,o}^{(k)}\hat{d}%
_{o}^{(k)}\rangle _{H_{j}}$. Note that, in addition to its
transforming the
microwave modes into optical modes, the EOM converter transforms the \emph{%
phase-sensitive} cross correlations, $\{\langle \hat{c}_{R}^{(k)}\hat{d}%
_{o}^{(k)}\rangle _{H_{1}}\}$, into \emph{phase-insensitive} cross
correlations, $\{\langle \hat{d}_{\eta
,o}^{(k)}\hat{d}_{o}^{(k)}\rangle
_{H_{1}}\}$, which can be measured---as shown in Fig.~\ref{mesurmentapp}%
---by mixing returned and retained modes on a 50--50 beam splitter whose
outputs are
\begin{equation}
\hat{a}_{\eta ,\pm }^{(k)}\equiv \frac{\hat{d}_{\eta ,o}^{(k)}\pm \hat{d}%
_{o}^{(k)}}{\sqrt{2}}.
\end{equation}%
These modes are then photodetected, yielding modal photon-counts
that are equivalent to measurements of the number operators
$\hat{N}_{\eta ,\pm }^{(k)}\equiv \hat{a}_{\eta ,\pm }^{(k)\dagger
}\hat{a}_{\eta ,\pm }^{(k)}$. Finally, the target
absence-or-presence decision is made by comparing the difference
of the two detectors' total photon counts \cite{footnote3}, which
is equivalent to the quantum measurement
\begin{equation}
\hat{N}_{\eta }=\sum_{k=1}^{M}\left( \hat{N}_{\eta
,+}^{(k)}-\hat{N}_{\eta ,-}^{(k)}\right) ,
\end{equation}%
with a threshold chosen to make the receiver's false-alarm
probability
\begin{equation}
P_{F}\equiv \Pr (\mbox{decide target present}\mid \mbox{target
absent}),
\end{equation}%
equals its miss probability,
\begin{equation}
P_{M}\equiv \Pr (\mbox{decide target absent}\mid \mbox{target
present}).
\end{equation}%
\begin{figure}[th]
\centering
\includegraphics[width=4.1in]{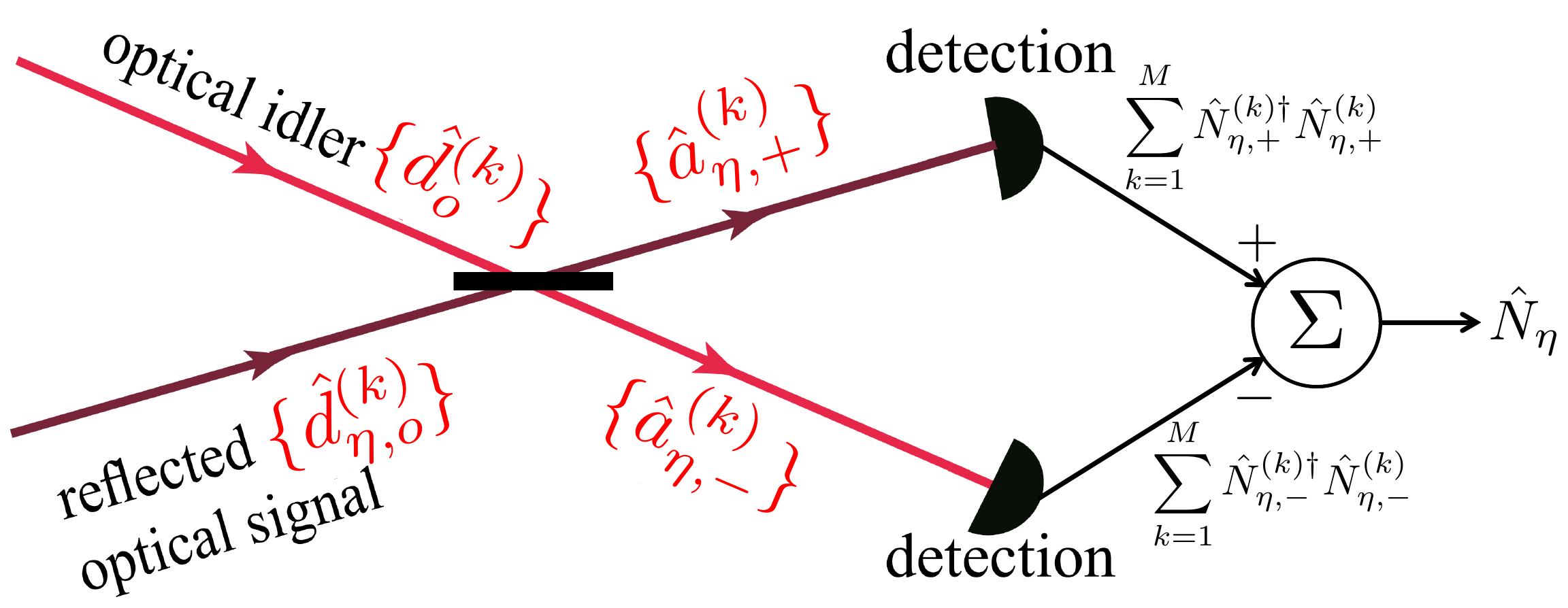}
\caption{Optical post-processing in the QI receiver. A 50-50 beam
splitter
mixes the wavelength-converted return modes $\{\hat{d}_{\protect\eta %
,o}^{(k)}\}$, from the receiver's EOM converter with the retained
idler modes, $\{\hat{d}_{o}^{(k)}\}$. The beam splitter's outputs
are detected,
yielding classical outcomes equivalent to the quantum measurements $%
\sum_{k=1}^{M}\hat{N}_{\protect\eta ,\pm }^{(k)}$, and the
difference of these outputs, equivalent to the quantum measurement
of $\hat{N}_{\eta}$, is used as the input to a threshold detector
(not shown) whose output is the target absence or presence
decision.} \label{mesurmentapp}
\end{figure}

Because QI employs an enormous number of mode pairs, and those
mode pairs
are iid under both hypotheses, the Central Limit Theorem implies that the $%
\hat{N}_\eta$ measurement yields a random variable that is
Gaussian, conditioned on target absence or target presence. It
follows that the QI receiver's error probability, for
equally-likely hypotheses, satisfies
\begin{equation}
P_\text{QI}^{(M)} = \frac{P_F + P_M}{2} = \frac{\text{erfc}\!\left(\sqrt{%
\text{SNR}_\text{QI}^{(M)}/8}\right)}{2},
\end{equation}
where the $M$-mode signal-to-noise ratio is
\begin{equation}
\text{SNR}_\text{QI}^{(M)} = \frac{4(\langle\hat{N}_\eta\rangle_{H_1}-%
\langle \hat{N}_\eta\rangle_{H_0})^2}{\left(\sqrt{\langle \Delta \hat{N}%
_\eta ^2\rangle_{H_0}} + \sqrt{\langle \Delta \hat{N}_\eta ^2\rangle_{H_1}}%
\right)^2},
\end{equation}
with $\langle\hat{N}_\eta\rangle_{H_j}$ and $\langle \Delta
\hat{N}_\eta ^2\rangle_{H_j}$, for $j = 0,1$, being the
conditional means and conditional variances of $\hat{N}_\eta$.

The iid nature of the $\{\hat{N}_{\eta ,+}^{(k)},\hat{N}_{\eta
,-}^{(k)}\}$ allow us to rewrite the $M$-mode SNR in terms of
single-mode moments, i.e.,
\begin{equation}
\text{SNR}_{\text{QI}}^{(M)}=\frac{4M[(\langle \hat{N}_{\eta
,+}\rangle _{H_{1}}-\langle \hat{N}_{\eta ,-}\rangle
_{H_{1}})-(\langle \hat{N}_{\eta ,+}\rangle _{H_{0}}-\langle
\hat{N}_{\eta ,-}\rangle _{H_{0}})]^{2}}{\left( \sqrt{\langle
(\Delta \hat{N}_{\eta ,+}-\Delta \hat{N}_{\eta ,-})^{2}\rangle
_{H_{0}}}+\sqrt{\langle (\Delta \hat{N}_{\eta ,+}-\Delta
\hat{N}_{\eta ,-})^{2}\rangle _{H_{1}}}\right) ^{2}},  \label{SNR}
\end{equation}%
where we have suppressed the superscript $^{(k)}$. The moments
needed to instantiate Eq.~(\ref{SNR}) are now easily found. In
particular, for the mean values we find that
\begin{subequations}
\begin{align}
\langle \hat{N}_{\eta ,\pm }\rangle _{H_{0}}& =|B|^{2}[(\bar{n}_{B}+\bar{n}_{%
\text{w}}^{T})/2+1]+|A_{o}|^{2}\bar{n}_{o}^{T}+|C_{o}|^{2}(\bar{n}%
_{b}^{T}+1), \\
\langle \hat{N}_{\eta ,\pm }\rangle _{H_{1}}& =\langle
\hat{N}_{\eta ,\pm
}\rangle _{H_{0}}+\eta |B|^{2}[|A_{\text{w}}|^{2}(\bar{n}_{\text{w}%
}^{T}+1)+|B|^{2}\bar{n}_{o}^{T}+|C_{\text{w}}|^{2}(\bar{n}_{b}^{T}+1)]/2
\notag \\
& \quad \pm \sqrt{\eta }\,\text{Re}[|B|^{2}A_{\text{w}}(\bar{n}_{\text{w}%
}^{T}+1)-|B|^{2}A_{o}\bar{n}_{o}^{T}+B^{\ast }C_{\text{w}}C_{o}(\bar{n}%
_{b}^{T}+1)]
\end{align}%
whence
\end{subequations}
\begin{subequations}
\begin{align}
\langle \hat{N}_{\eta ,+}\rangle _{H_{0}}-\langle \hat{N}_{\eta
,-}\rangle
_{H_{0}}& =0, \\
\langle \hat{N}_{\eta ,+}\rangle _{H_{1}}-\langle \hat{N}_{\eta
,-}\rangle
_{H_{1}}& =2\sqrt{\eta }\,\text{Re}[|B|^{2}A_{\text{w}}(\bar{n}_{\text{w}%
}^{T}+1)-|B|^{2}A_{\text{o}}\bar{n}_{\text{o}}^{T}+B^{\ast }C_{\text{w}%
}C_{o}(\bar{n}_{b}^{T}+1)],
\end{align}%
\begin{figure}[th]
\centering
\includegraphics[width=3.1in]{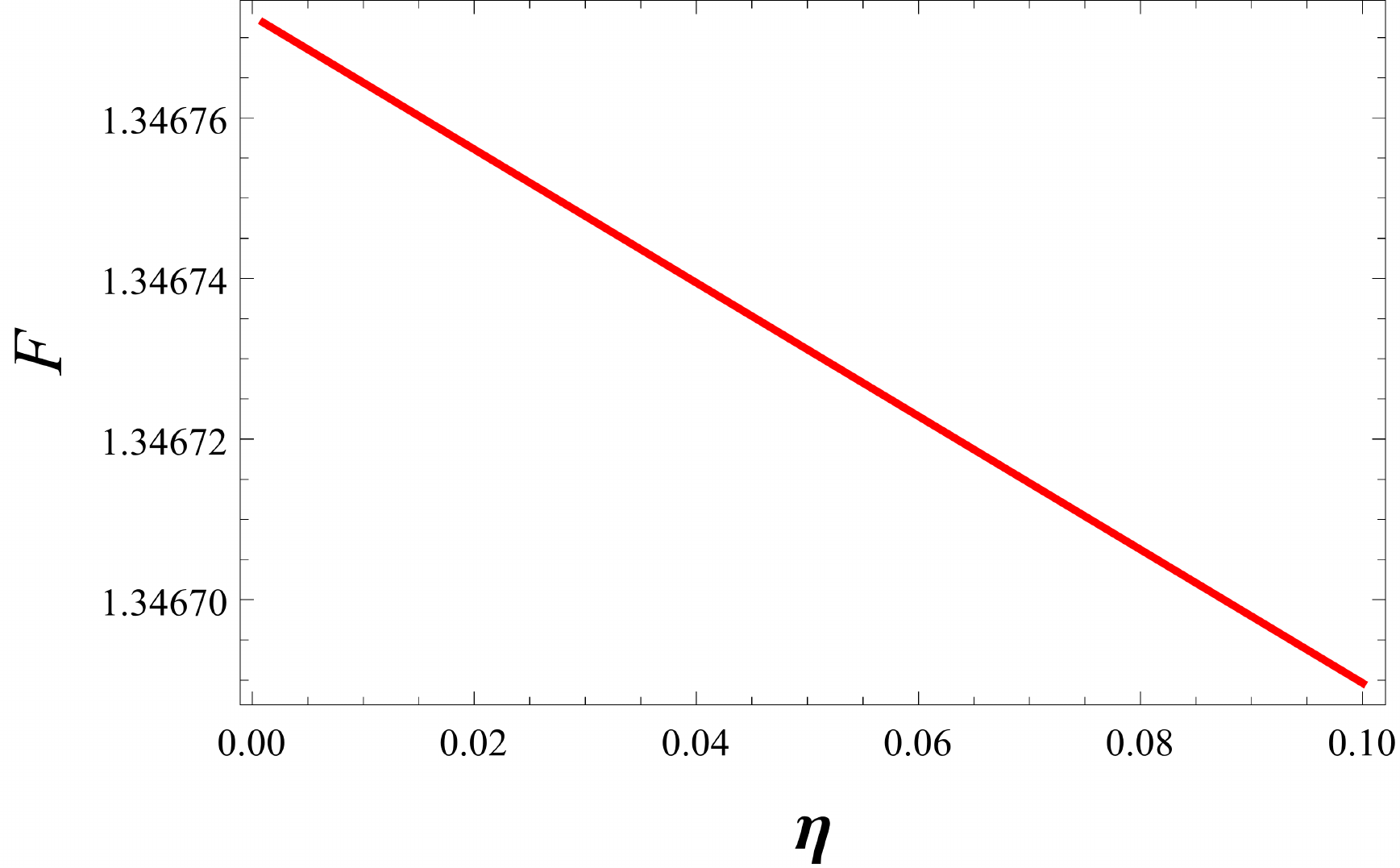}
\caption{The dependence of the QI-advantage figure of merit $\mathcal{F}%
\equiv \text{SNR}_{\text{QI}}^{(M)}/\text{SNR}_{\text{coh}}^{(M)}$ on $%
\protect\eta $. See Fig. 2 of the main paper for the other
parameter values.} \label{etaPIC}
\end{figure}
For the variances we get~\cite{saikatAPP}
\end{subequations}
\begin{equation}
\langle (\Delta \hat{N}_{\eta ,+}-\Delta \hat{N}_{\eta
,-})^{2}\rangle _{H_{j}}=\langle \hat{N}_{\eta ,+}\rangle
_{H_{j}}(\langle \hat{N}_{\eta ,+}\rangle _{H_{j}}+1)+\langle
\hat{N}_{\eta ,-}\rangle _{H_{j}}(\langle
\hat{N}_{\eta ,-}\rangle _{H_{j}}+1)-(\langle \hat{d}_{\eta ,o}^{\dagger }%
\hat{d}_{\eta ,o}\rangle _{H_{j}}-\langle \hat{d}_{o}^{\dagger }\hat{d}%
_{o}\rangle )^{2}/2,
\end{equation}%
for $j=0,1$, where
\begin{equation}
\langle \hat{d}_{\eta ,o}^{\dagger }\hat{d}_{\eta ,o}\rangle
_{H_{j}}=\left\{
\begin{array}{ll}
|B|^{2}(\bar{n}_{B}+1)+|A_{o}|^{2}\bar{n}_{o}^{T}+|C_{o}|^{2}(\bar{n}%
_{b}^{T}+1), & \mbox{ for $j = 0$} \\[0.05in]
\langle \hat{d}_{\eta ,o}^{\dagger }\hat{d}_{\eta ,o}\rangle
_{H_{0}}+\eta
|B|^{2}[|A_{\text{w}}|^{2}(\bar{n}_{\text{w}}^{T}+1)+|B|^{2}\bar{n}%
_{o}^{T}+|C_{\text{w}}|^{2}(\bar{n}_{b}^{T}+1)], & \mbox{for $j = 1$.}%
\end{array}%
\right.
\end{equation}

It is interesting to see how the transmitter-to-target-to-receiver
transmissivity, $\eta $, affects QI's performance advantage over
coherent-state operation. Figure~\ref{etaPIC} shows the dependence
of the
QI-advantage figure of merit $\mathcal{F}\equiv \text{SNR}_{\text{QI}}^{(M)}/%
\text{SNR}_{\text{coh}}^{(M)}$ for $10^{-3}\leq \eta \leq
10^{-1}$. We have restricted our attention here to the low-$\eta $
regime, because we follow the lead of~\cite{pirandolaAPP} in
assuming that the target's presence has no passive signature. A
passive signature---a lower received background level in the
presence of the target---would permit the target to be detected
without transmitting any microwave radiation. We preclude that
possibility
by assuming that the noise mode $\hat{c}_{B}$ has average photon number $%
\bar{n}_{B}$ when the target is absent and average photon number $\bar{n}%
_{B}/(1-\eta )$ when the target is present. Unless $\eta \ll 1$,
so that these two average photon numbers are nearly the same, this
assumption is physically unreasonable.


\begin{thebibliography}{99}
\bibitem{bandwidth} The common bandwidth of the entangled microwave and
optical fields is $W_m = \gamma_M(\Gamma_\text{w}-\Gamma_o + 1)$, in terms
of the resonator's damping rate and the cooperativity parameters.
\bibitem{Stability} The system is stable if the Routh-Hurwitz criterion is
satisfied. For $\Gamma _{i}\gg 0$, this criterion reduces to the
following necessary and sufficient condition~\cite{Wang}:
\begin{equation*}
\kappa _{\text{w}}\Gamma _{\text{w}}-\kappa _{o}\Gamma _{o}>\tilde{K}\,\text{max}\left\{ \kappa _{\text{w}}-\kappa _{o},\frac{\kappa _{o}^{2}-\kappa _{\text{w}}^{2}}{2\gamma _{M}+\kappa _{\text{w}%
}+\kappa _{o}}\right\} ,
\end{equation*}%
where $\tilde{K}=\Gamma _{\text{w}}(1+\kappa
_{o}/\kappa_{\text{w}})^{-1}+\Gamma _{o}(1+\kappa
_{\text{w}}/\kappa _{o})^{-1}$.
\end{thebibliography}
\begin{thebibliography}{9999}
\bibitem{footnote1} The classical bound arises from the requirement that $%
\langle |x d_\text{w} + d_o^*|^2\rangle \ge 0$ for classical,
complex-valued random variables $d_\text{w}$ and $d_o$, where
$\langle \cdot\rangle$ denotes classical ensemble average and $x$
is an arbitrary real-valued parameter.
\bibitem{hx} Note that the expression of the entropic function $h(x)$ is
that for vacuum noise equal to $1/2$. Our notation is different
from that of Ref.~\cite{RMP}, where the vacuum noise is equal to
$1$.
\bibitem{footnote3} For continuous-wave operation with a $t_m$-sec-long
measurement interval, the $\hat{N}_\eta$ measurement is realized
by the photon-count difference between the two detectors over that
measurement interval.
\end{thebibliography}
\end{document}